%% file: main.tex
\documentclass[a4paper,11pt]{article}
\usepackage{jheppub} 
\usepackage{lineno}
\usepackage{multirow}
\usepackage{bm}
\usepackage{tensor}
\usepackage{longtable}
\usepackage{xcolor}
 \usepackage[normalem]{ulem}

\toccontinuoustrue

\usepackage{amsmath}
\newcommand*{\defeq}{\stackrel{\text{def}}{=}}
\newcommand{\intwK}{\kappa^{-1}\!\int \!\! d \mathbf{x} \,}
\newcommand{\intnK}{\int \!\! d \mathbf{x} \,}

\title{\boldmath Anomaly freedom in effective Loop Quantum Cosmology: pedagogical summary and generalized holonomy corrections}

\author{Maxime De Sousa,}
\author{Aurélien Barrau,}
\author{Killian Martineau}
\affiliation{Laboratoire de Physique Subatomique et de Cosmologie, Universit\'e Grenoble-Alpes, CNRS/IN2P3\\
53, avenue des Martyrs, 38026 Grenoble cedex, France}

\emailAdd{maxime.desousa@lpsc.in2p3.fr}

\compress
\abstract{The issue of consistency is crucial in quantum gravity. It has recently been intensively addressed for effective symmetry-reduced models. In this article, we exhaustively study the anomaly freedom of  effective loop quantum cosmology with generalized holonomy corrections, considering loop correction of the constraints at the perturbative order. 
We pedagogically explain why, although the holonomy correction -- including the details of the chosen scheme -- applied on the background part of the constraints is crucial, it becomes irrelevant when implemented on perturbative expansions, in the sense that all consequences are ``absorbed" in the counter-terms used for the regularization.
The possibility of closing the algebra of constraints without counter-terms is also studied. It is argued that, although enforcing a first-class algebra is a strong requirement, this can be achieved in several different ways, often overlooked, which generates ambiguities on the restriction of the form of the generalized holonomy correction. Those ambiguities are examined in details, leading to the conclusion that the consistency of the effective theory for cosmological perturbations, especially when considering scalar modes, cannot be achieved without counter-terms. We also take the opportunity of this work to clarify, as much as possible, all the required steps so that future works have a clear material at disposal.
In particular, a highly detailed calculation of all the brackets is provided, emphasizing  the (usually implicit) assumptions, hypotheses and manipulations required to ensure the closure of the algebra. Prospects for future works are underlined.}

\begin{document}
\maketitle
\allowdisplaybreaks

\section{Introduction}

\label{sec:intro}
    \input{texts/introduction}

\section{Effective Loop Quantum Cosmology}
    \input{texts/lqc/intro_lqc}
\subsection{Basic variables and effective quantum corrections}
\subsubsection{Background variables}
    \input{texts/lqc/variables}

\subsubsection{Generalized holonomy corrections}
    \input{texts/lqc/ghc}

\subsection{Perturbations}
    \input{texts/lqc/perturbations}
    
\subsection{Gravitational constraints}
    \label{sec:constraints}
    \input{texts/constraints/intro}

\subsubsection{Gauss constraints}
    \input{texts/constraints/gaussian}

\subsubsection{Diffeomorphism constraints}
    \label{sec:geo_diffeo}
    \input{texts/constraints/geo_diffeo}

\subsubsection{Hamiltonian constraint}
    \input{texts/constraints/geo_scalar}

\subsection{Gravity coupled to a scalar field: constraints}
    \input{texts/constraints/intro_scalar_field}

\subsubsection{Diffeomorphism constraints}
    \input{texts/constraints/matter_diffeo}

\subsubsection{Hamiltonian constraint}
    \input{texts/constraints/matter_scalar}

\subsection{Algebra of constraints}
    \label{sec:algebra}
    \input{texts/constraints/algebra}
    
\section{Algebra computations}
\label{sec:AlgComputations}
\subsection{Bracket $\left\{ \mathbb{G}, \mathbb{G} \right\}$}

\input{texts/brackets/G-G}

\subsection{Bracket $\left\{ \mathbb{D}, \mathbb{D} \right\}$}
    \input{texts/brackets/D-D}

\subsection{Bracket $\left\{ \mathbb{D}, \mathbb{G} \right\}$}
    \input{texts/brackets/D-G}
    
\subsection{Bracket $\left\{ \mathbb{H}, \mathbb{G} \right\}$}
    \input{texts/brackets/H-G}

\subsection{Bracket $\left\{ \mathbb{H}, \mathbb{D} \right\}$}
    \input{texts/brackets/H-D}

\subsection{Bracket $\left\{ \mathbb{H}, \mathbb{H} \right\}$}
    \input{texts/brackets/H-H}
    
\section{Solutions to close the algebra of constraints}
\subsection{Closure via a deformation of the constraints}
    \label{sec:closure_ct}
    \input{texts/solution_algebra/solution_ct}
\subsection{Closure via a restriction on the GHCs}
    \label{sec:closure_ghc}
    \input{texts/solution_algebra/solution_restriction_ghc}

\section{Conclusion and prospects}
\label{sec:conclusion}
    \input{texts/conclusion}

\newpage
\appendix
\section{Anomalies: Summary of computations}
    \input{texts/appendices/summary_anomalies}
\newpage
\section{Anomalies: Table of correspondence}
    \input{texts/appendices/anomalies_table_correspondance}

\bibliographystyle{JHEP}
\bibliography{biblio.bib}

\end{document}

%% file: texts/introduction.tex
The quest for a quantum theory of gravity is one of the most difficult tasks for theoretical physics (see \cite{Kiefer:2005uk,Oriti:2009zz} for reviews of currently followed paths). No consensual approach exists today but different models are being considered. One way to compare them is obviously to carefully investigate their respective observational predictions (see, e.g., \cite{Barrau:2017tcd} and references therein). This is the most obvious approach but it is extremely difficult in practice as the Planck scale is 15 orders of magnitude beyond what is currently probed with colliders (see \cite{Rovelli:2021thx} for some thoughts on the subject). Another key aspects is the one of {\it consistency}. Although all serious models are, by construction, consistent at first sight, a careful analysis can reveal severe flaws that were not {\it a priori} obvious. This is one of the central motivation for the so-called swampland program in string theory (see \cite{Palti:2019pca} for a review): some theories believed to lie in the landscape cannot be pushed to high energies in the presence of gravity. This article focuses on a different aspect, namely the closure of the algebra of constraints. It is a very important consistency condition as it ensures that the vectors of evolution are parallel to the sub-manifolds of constraints. Otherwise stated, this guaranties that, starting from a physical solution and using the dynamical laws of the theory, one still gets a physical solution.\\

When the equations of gravity -- especially those describing the evolution of cosmological perturbations -- are quantum corrected at the effective level it is difficult to determine whether the subtle consistency conditions embedded in the first-class nature of the algebra of constraints are still satisfied \cite{Barrau:2014maa}. Many modern approaches to canonical quantum general relativity \cite{Thiemann:2007pyv} put the emphasis on background-independence as being the conceptual core of Einstein's gravity. Those frameworks cannot take advantage of usual covariance criteria as space-time itself should, in a way, emerge from solutions to their dynamical equations.\\
Generally speaking, gauge fixing before quantization is dangerous. Observables can be significantly different from what would have been obtained if the theory had been quantized without fixing the gauge. In gravity, dynamics is part of the gauge and one should be very careful not to fix the gauge according to transformations that have to be subsequently modified \cite{Bojowald:2008gz}. The issue of possible anomalies in the Poisson brackets between constraints in a central one in this framework.\\

In the last decade, a tremendous amount of work has been devoted to the question of anomaly freedom in effective loop quantum gravity (LQG), see, e.g. \cite{Rovelli:2011eq,Rovelli:2014ssa,Ashtekar:2017yom,Ashtekar:2021kfp} for pedagogical introductions to LQG. Notably, great progresses were recently made in the black hole sector \cite{BenAchour:2018khr,Arruga:2019kyd,Alonso-Bardaji:2023niu, Alonso-Bardaji:2022ear, Alonso-Bardaji:2020rxb, Alonso-Bardaji:2021tvy,Belfaqih:2024vfk}. The present work focuses on cosmological aspects as it seems to us that a pedagogical and rigorous derivation of all the relevant steps was missing. Currently, anyone interested in this approach has to reinvent all the machinery, which is far from being light or trivial. The first goal of this article is therefore to explain the subtleties and technicalities -- some of them being sometimes erroneously neglected in the literature -- so that future works can rely on a clear basic material where hypotheses are made explicit. New results regarding the use of generalized holonomy corrections (GHC), which are currently being intensively considered (in particular, but not only, because of points made in \cite{Perez:2005fn,Vandersloot:2005kh,BenAchour:2016ajk,Amadei:2022zwp}), are derived. We point out that although the holonomy correction -- including its detailed expression -- plays a crucial role when implemented at the background level (see, e.g., \cite{Han:2017wmt,Renevey:2021tmh,DeSousa:2022rep} for the importance of the GHC), it brings no new effect at all when implemented in the perturbative expansion of the constraints\footnote{This is true when considering only, as done in all studies including this one, a modification of the background variable -- even in the perturbed constraints. The question of what happens when correcting the perturbed variables themselves is still opened.}. We also underline that the closure of the algebra cannot be achieved with a specific expression of the generalized holonomy correction, without the use of counter-terms.\\

The first part of this work review the basics of loop quantum cosmology for the unfamiliar reader. Then, the detailed calculation of all the Poisson brackets between constraints is performed. Finally, different ways to close the algebra are considered. Promising new directions for future works are also pointed out. Throughout all the work, assumptions that could be relaxed in future studies are underlined.\\

%% file: texts/lqc/intro_lqc.tex
Loop Quantum Cosmology (LQC) is a framework for the description of the very early universe, mimicks quantization techniques of Loop Quantum Gravity (LQG) in the cosmological sector. It has rapidly become a valuable testing ground for addressing and evading some of the technical challenges of full LQG, mostly because the LQC model is exactly solvable \cite{Ashtekar:2011ni}. Notably, in the last decades, significant insights have emerged regarding the history of the early universe, such as the fact that the matter density operator $\hat{\rho}$ has an upper bound when LQC is coupled to a scalar field $\phi$, which indicates a resolution of the ``big-bang" singularity. Additionally, the effective constructions in LQC successfully capture the dynamics of semi-classical states~\cite{Ashtekar:2006wn}. In order to make this article self-contained and set the notations, a brief introduction to LQC is provided in the next subsection. For nice and more detailed reviews of LQC, we refer to \cite{Ashtekar:2011ni, Bojowald:2005epg, Agullo:2023rqq}.\\

Several ways to study cosmological perturbations in the LQG paradigm do exist. Among interesting paths, are the so called ``dressed metric" approach \cite{Agullo:2012sh,Agullo:2012fc,Agullo:2013ai} and the (related) ``hybrid quantization" scheme \cite{Garay:2010sk,Fernandez-Mendez:2012poe}. Comparisons between existing models can be found, e.g., in \cite{Bolliet:2015bka,Ashtekar:2015dja,Wilson-Ewing:2016yan,Li:2022evi}. This work shall focus only on the ``deformed algebra" approach which, in a way, might capture less quantum effects but puts the emphasis on consistency, which is certainly a necessary requirement. Importantly, in a sense, it also goes beyond the LQG/LQC models and constitutes a relevant framework to investigate a wide class of possible deformations of general relativity \cite{Bojowald:2024fjk, Bojowald:2024naz, Cuttell:2019fgl, Arruga:2019kyd}. 

%% file: texts/lqc/variables.tex
The line element of a space-time $\mathcal{M}$ parameterized by $\mathcal{M}=\mathbb{R} \times \Sigma$, where $\Sigma$ corresponds to spatial slices, can be given by the ADM formalism \cite{Arnowitt:1959ah}
\begin{equation}
    ds^2 = - N^2 d\eta^2 + q_{ab} \bigl( N^a d\eta + dx^a\bigr)\bigl( N^b d\eta + dx^b\bigr), \label{eq:LineElemADM}
\end{equation}
in which $N$ is the lapse function, $N^a$ is the shift vector and $q_{ab}$ is the induced spatial metric on $\Sigma$. Throughout all this work, we consider a flat FLRW space-time as the background for cosmological perturbations theory, whose line element is given by \cite{Weinberg:2008zzc}:
\begin{equation}
    ds^2 = a^2(\eta)\bigl( -d\eta^2 + \delta_{ab} dx^a dx^b\bigr). \label{eq:LineElemFLRW}
\end{equation}
Direct comparison of Eq. (\ref{eq:LineElemADM}) and Eq. (\ref{eq:LineElemFLRW}) leads to the following identifications:
\begin{equation}
    q_{ab}=a^2(\eta)\delta_{ab}, \quad N=a(\eta), \quad N^a=0.
\end{equation}
The LQG canonical quantization program is based on the so-called Ashtekar's variables \cite{Ashtekar:1987gu,BarberoG:1994eia}. In this setup, the densitized triads $E^a_i$ enable the building of the 3-metric $q_{ab}$, such that $E^a_i E^b_j \delta^{ij} \defeq \bigl| \text{det} \, q \bigr| q^{ab}$. At the background level, the densitized triads are given by \cite{Bojowald:2005epg}:
\begin{equation}
    E^a_i = a^2(\eta)\delta^a_i \defeq \mathfrak{p}(\eta) \delta^a_i. \label{eq:TriadstFLRWReduced}
\end{equation}
Their canonically conjugate variables $K^i_a$ are built from the extrinsic curvature tensor
\begin{equation}
    K_{ab} \defeq \bigl( 2 N\bigr)^{-1} \bigl( \partial_\eta q_{ab} - 2 D_{(a}N_{b)} \bigr)= \frac{\partial_\eta \mathfrak{p}}{2 \sqrt{\mathfrak{p}}} \delta_{ab},
\end{equation}
where $D$ denotes the spatial covariant derivative of $q_{ab}$. From the previous expression, one can construct \cite{Bojowald:2005epg},
\begin{equation}
    K^i_a = \frac{E^{bi}}{\sqrt{\bigl| \text{det} E \bigr|}}K_{ab} = \frac{\partial_\eta \mathfrak{p}}{2 \mathfrak{p}} \delta^i_a \defeq \mathfrak{c}(\eta)\delta^i_a,
\end{equation}
such that $K^i_a$ and $E^a_i$ are canonically conjugate. They fulfill the relations
\begin{equation}
    \biggl\{K^i_a(\mathbf{x}), E^b_j(\mathbf{y}) \biggr\} \defeq \kappa \! \int \! \! d\mathbf{z} \biggl( \frac{\delta K^i_a(\mathbf{x})}{\delta K^l_c(\mathbf{z})} \frac{\delta E^b_j(\mathbf{y})}{\delta E^c_l(\mathbf{z})} \biggr)= \kappa \delta^i_j \delta^b_a \delta(\mathbf{x}-\mathbf{y}),
    \label{Poisson bracket K-E}
\end{equation}
with $\kappa=8 \pi G$, from which the Poisson bracket between $\frak{p}$ and $\frak{c}$ can be obtained:
\begin{equation}
    \bigl\{\frak{c}, \frak{p}\bigr\} =\frac{\kappa}{3} \mathcal{V}^{-1}.
\end{equation}
In the previous expression $\mathcal{V}$ is defined as the volume of a fiducial cell used to regularize the integrals appearing in the Poisson brackets of the background variables in Eq. (\ref{Poisson bracket K-E}).\\

As usual, canonical transformations can be usefully performed in the Hamiltonian framework. In particular, the use can me made of the so-called Ashtekar's connection, defined as~\cite{Thiemann:2007pyv}: 
\begin{equation}
    A^i_a = \Gamma^i_a + \gamma K^i_a,
\end{equation}
where $\Gamma^i_a$ is the $\frak{so}(3)$-valued spin connection compatible with the triads, and $\gamma$ is the Barbero-Immirzi (free) parameter. This canonically conjugate, equivalent, phase-space is now parameterized via the densitized triads $E^a_i$ and the Ashtekar's connection $A^i_a$ verifying the structure
\begin{equation}
     \biggl\{A^i_a(\mathbf{x}), E^b_j(\mathbf{y}) \biggr\} = \gamma \kappa \delta^i_j \delta^b_a \delta(\mathbf{x}-\mathbf{y}).
\end{equation}
In addition to the gravitational sector, the energy content is assumed to be a minimally coupled scalar field $\phi$, whose action is given by:
\begin{equation}
S_{\phi} = \int \! \! d^4x \sqrt{|\text{det} g|} \left( \frac{1}{2} \nabla_\mu \phi \nabla^\mu \phi - V[\phi] \right),
\end{equation}
where $V[\phi]$ is the potential of the scalar field. \\

In a symplectic framework, the canonical variable $\phi$ and its conjugate momentum $\pi$ satisfy the relation
\begin{equation}
     \biggl\{\phi(\mathbf{x}), \pi(\mathbf{y}) \biggr\}= \delta(\mathbf{x}-\mathbf{y}).
\end{equation}
The background values $\bm{\phi}$ and $\bm{\pi}$ of those fields satisfy the standard Klein-Gordon dynamics. Their Poisson bracket is
\begin{equation}
     \bigl\{\bm{\phi}, \bm{\pi} \bigr\}= \mathcal{V}^{-1}.
\end{equation}
The evolution of the canonical variables is driven by constraints, which will be detailed in the following.

%% file: texts/lqc/ghc.tex
As stated above, in canonical LQG, geometry is described by the Ashtekar connection $A^i_a$ and the densitized triads $E_i^a$ instead of the spatial metric $q^{ab}$ and its associated momentum $\Pi_{ab}$ \cite{Rovelli:2011eq,Rovelli:2014ssa,Ashtekar:2017yom,Ashtekar:2021kfp}. This is a step towards background independence, as these fields can be smeared without referring to the metric. Specifically, the Ashtekar's connection is smeared along closed curves to produce holonomies, while the densitized triads are naturally integrated over the associated surfaces \cite{Ashtekar:2017yom,Ashtekar:2021kfp}. The quantization ``à la Dirac" of the resulting holonomies and fluxes leads to discrete spectra for geometrical operators, such as the area~\cite{Rovelli:1998gg, Rovelli:1994ge} or the volume \cite{Rovelli:1994ge, Ashtekar:1997fb}. Specifically, the regularization of the constraints (see Sec. \ref{sec:constraints}) with respect to holonomies significantly impacts the dynamics~\cite{Ashtekar:2006rx}. The deep reasons for the ``periodization" of the connection variable is that there is no natural quantum operator associated with the Ashtekar connection (see, e.g., \cite{Gambini:2011zz}). At the effective level, the use of holonomies instead of the connection itself modifies the expression of the curvature tensor.
This modification is described by the so-called ``holonomy correction", which consists in the  replacement \cite{Ashtekar:2011ni}
\begin{equation}
	\frak{c} \longrightarrow \frac{\sin(\delta\frak{c})}{\delta}
 \label{holonomy correction equation}
\end{equation}
in the curvature tensor. The explicit expression of $\delta \defeq \delta(\frak{p})$ depends on the scheme under consideration~\cite{Ashtekar:2011ni}.\\

Nonetheless, the quantization procedure of the Hamiltonian constraint is subject to many ambiguities~\cite{Thiemann:2006cf} and, as a consequence, so is the resulting correction \cite{Perez:2005fn}. A tremendous amount of work has been devoted to those ambiguities in the context of LQC (see, e.g., \cite{Liegener:2019dzj, Vandersloot:2005kh, BenAchour:2016ajk, Hrycyna:2008yu, Yang:2009fp}). In this work, to remain as generic as possible, we follow \cite{Amadei:2022zwp, Han:2017wmt} and introduce a so-called Generalized Holonomy Correction (GHC), which corresponds to the replacement:
\begin{equation}
    \frak{c} \longrightarrow g(\frak{c}, \frak{p}).
    \label{eq:ghc}
\end{equation}
There is \textit{a priori} no restriction on $g$ as long as $g\rightarrow \frak{c}$ in the low curvature regime.\\ 

Throughout this work, only the background variable $\frak{c}$ is holonomy-corrected. However, we still introduce a second correction:
\begin{equation}
    \frak{c} \longrightarrow \tilde{g}(\frak{c}, \frak{p}),
\end{equation}
that will be applied to the perturbed constraints. Relying on two distinct corrections will help discriminating between the effects due to the use of a GHC at the background level and those arising from the use of the GHC for the perturbations. \\

As we shall see in the following sections, this second correction appears frequently in various calculations, including anomalies and counter-terms. However, it will be shown that the resulting Hamiltonian depends functionally only on $g$, thereby leading to the conclusion that physical observables remain unaffected by the introduction of a GHC in the perturbed constraints. This shows that including holonomy corrections in the background part of the Hamiltonian is sufficient. In addition, in the last section, we will explain why, without counter-terms, those GHCs are actually restricted to the usual GR expressions.  \\

It is important to stress that, in this work, we follow the usual approach in which the holonomy correction appears in the perturbations only through modifications of the background extrinsic curvature $\frak{c}$. An important step further would be to perform a rigorous treatment of the perturbative expansion of the holonomy, based on the perturbed connection. This goes beyond the scope of this article but it should be considered in future works.

%% file: texts/lqc/perturbations.tex
The study of cosmological perturbations is crucial, mainly for two reasons. Firstly, deriving the cosmological power spectrum is the key ingredient for comparison with observations.
Secondly, as it will be discussed later, the gauge freedoms related to the Gauss constraint $\mathbb{G}$ and to the (spatial) diffeomorphism constraint $\mathbb{D}$ are fixed for homogeneous models -- ensuring consistency (see sect. \ref{sec:algebra}). 
Perturbations can be considered as test fields  revealing the underlying space-time structure \cite{Teitelboim:1972vw}, e.g. its signature, that cannot be seen at the background level.
The dynamics of perturbations has to be calculated within a consistent framework.
The inhomogeneities of the gravitational phase-space are parametrized by
\begin{equation}
    E^a_i = \frak{p} \delta^a_i + \delta E^a_i , \quad \text{ and } \quad K^i_a = \frak{c} \delta^i_a + \delta K^i_a,
\end{equation}

and those of the matter phase-space by

\begin{equation}
    \phi = \bm{\phi} + \delta \phi , \quad \text{ and } \quad \pi = \bm{\pi} + \delta \pi.
\end{equation}

Moreover, to describe perturbations of all the components of the 4-metric $g_{\mu\nu}$, perturbations around the mean value of the lapse and shift functions must also be considered. Those perturbations are parametrized by

\begin{equation}
    N = \mathbf{N} + \delta N, \quad \text{ and } \quad N^a = \mathbf{N}^a + \delta N^a.
\end{equation}
In this paper, our focus lies on the evolution equations of cosmological perturbations at the \textit{linear} order.
The equations of motion of a phase space quantity $f$ governed by a Hamiltonian $H$ being given by $\partial_t f = \left\{f, \, H\right\}$, it is necessary to consider the perturbation of the Hamiltonian $H$ at the second order. This is essentially implied by a ``loss of perturbation order" when the Poisson bracket is applied to the perturbed phase space. The perturbation of the Hamiltonian being pushed to the second order, a question might then arise: why are second order perturbations of the canonical variables neglected? The reason lies in the fact that, \textit{within the Hamiltonian framework at the second order}, the \textit{second order} perturbations of canonical variables has no consequence on the \textit{linear} dynamics of perturbations.\footnote{
This trivially results from the very definition of the Poisson brackets. As an example, let us consider the \textit{linear} dynamic of the densitized triads $E^a_i$. Given a Hamiltonian $H$, from the definition of equations of motion, one has,
\begin{equation}
    \partial_t E^a_i =  \biggl\{ E^a_i(\mathbf{x}), H(\mathbf{y}) \biggr\} \defeq \kappa \! \int \! \! d\mathbf{z} \,\delta\left(\mathbf{x}-\mathbf{z}\right) \delta^a_b \delta^j_i  \frac{\delta \, H(\mathbf{y})}{\delta K^j_b(\mathbf{z})}.
\end{equation}
Thus, to have $\partial_t E^a_i$ at the linear order, $H$ has to be at the second order in perturbations and include $\delta K$ (otherwise it would be irrelevant for the dynamics of $E^a_i$). The same reasoning follows for each perturbed canonical variable.} The first significant contributions would be at the third order of perturbations to the Hamiltonian, corresponding to corrections with respect to the dynamics addressed in this paper.

%% file: texts/constraints/intro.tex
As previously discussed, this work is carried out at the effective level within the classical Hamiltonian formulation, expressed in Ashtekar's variables, of holonomy-corrected GR. As underlined by Dirac \cite{Dirac:1958sc}, the constraints play a key role in deriving physical quantities for gauge theories. Specifically, constraints 
\begin{enumerate}
\item restrict the fields to the hypersurface on which the constraints vanish,
\item generate gauge transformations for the fields,
\item provide the equations of motion of the fields.
\end{enumerate}

In this section, we briefly describe each constraint, focusing on the implementation of holonomy corrections.

%% file: texts/constraints/gaussian.tex
Due to the choice of working with Ashtekar's variables, which extend the phase space of GR, a constraint known as the Gaussian constraint do arise and is expressed as~\cite{Han:2005km}:
\begin{equation}
    \mathbb{G} \defeq \mathbb{G}[\Lambda^i] = \bigl( \kappa \gamma \bigr)^{-1}\!\int \! \! d\mathbf{x} \, \Lambda^i \mathcal{G}_i=\bigl( \kappa \gamma \bigr)^{-1}\!\int \! \! d\mathbf{x} \, \Lambda^i \biggl[ \partial_a E^a_i + \epsilon^l_{\, ik} A^k_a E^a_l \biggr],
\end{equation}
where $\mathcal{G}_i$ are the so-called Gaussian constraints densities and $\epsilon^l_{\, ik}$ is the Levi-Civita symbol. Gauge rotations in the internal space of phase-space functions are parameterized by the Lagrange multipliers $\Lambda^i$, such that the gauge transformations are given by~$\delta_\Lambda(f)~\defeq~\left\{f, \, \mathbb{G}[\Lambda^i]\right\}$. Gauge transformations of canonical variables, like the densitized triads $E^a_i$, can therefore be calculated as:
\begin{equation}
    \delta_\Lambda(E^a_i) \defeq \biggl\{E^a_i(\mathbf{x}), \, \mathbb{G}[\Lambda^i]\biggr\}=\gamma\!\int \! \! d\mathbf{z}\,\delta\left(\mathbf{x}-\mathbf{z}\right) \delta^a_b \delta^j_i  \frac{\delta \, \mathbb{G}[\Lambda^i]}{\delta A^j_b(\mathbf{z})}=\kappa^{-1}\epsilon_{ij}^{\, \, \,k}\Lambda^jE^a_k.
\end{equation}

After performing some identifications with the symmetry-reduced triads Eq. (\ref{eq:TriadstFLRWReduced}), it is easy to see that no gauge freedom, with respect to the Gauss constraint, is left if cosmological symmetries are applied. Consequently, there is a single way to construct the Gauss constraint at the second order in perturbations, which is,
\begin{equation}
    \mathbb{G}=\bigl( \kappa \gamma \bigr)^{-1}\!\int \! \! d\mathbf{x} \, \delta\Lambda^i \mathcal{G}^{(1)}_i, \label{eq:GaussExpansionDef}
\end{equation}
with,
\begin{equation}
    \mathcal{G}_i^{(1)}= \gamma \biggl[ \mathfrak{p} \, \epsilon_{ij}^a \delta K^j_a + \mathfrak{c} \, \epsilon_{ia}^j \delta E^a_j \biggr],
\end{equation}
where $X^{(n)}$ stands the perturbed expression of the quantity $X$ at the $n$-th order. \\

In LQG, the kernel of the Gauss constraints related operator is easily obtained with the spin-network decomposition (for details, see \cite{Han:2005km}). Schematically, it consists in demanding invariance of cylindrical functions under the transformations generated by classical Gauss constraints \cite{Bodendorfer:2016uat}. This ensures that the theory which is quantized is indeed GR, and not an extended gravity theory. One can thus study the canonical transformations and their structure modulo Gauss constraints: no quantum corrections are expected to arise here at the effective level. In particular, no GHC is to be implemented in the Gauss constraints, even though $\mathcal{G}_i^{(1)}$ depends on the reduced curvature~$\frak{c}$.

%% file: texts/constraints/geo_diffeo.tex
Diffeomorphism invariance is at the heart of GR. It reflects the independence of physical laws with respect to choices of coordinate systems.
In a canonical setup, the invariance under (spatial) diffeomorphisms is guaranteed if the so-called diffeomorphism constraints are satisfied. Modulo Gauss constraints, they are given by \cite{Han:2005km}:
\begin{equation}
    \mathbb{D}_{\mathfrak{g}}\defeq \mathbb{D}_{\mathfrak{g}}[N^a] = \bigl( \kappa \gamma \bigr)^{-1}\!\int \! \! d\mathbf{x} \, N^a \mathcal{D}_{a}^{\frak{g}}=\bigl( \kappa \gamma \bigr)^{-1}\!\int \! \! d\mathbf{x} \, N^a\biggl[ \left( \partial_a A^j_b - \partial_b A^j_a \right) E^b_j + A^j_a \partial_b E^b_j \biggr],
\end{equation}
where $\mathcal{D}_{a}^{\frak{g}}$ are the (geometrical) diffeomorphism constraints densities.\\

In the flat FLRW spacetime, as described by Eq. (\ref{eq:LineElemFLRW}), the lapse vector $N^a$ vanishes at the background level and the smeared (geometrical) diffeomorphism constraint can be non-ambiguously expressed at the second order in the perturbed canonical variables. It is given by:
\begin{equation}
    \mathbb{D}_{\mathfrak{g}} = \bigl( \kappa \gamma \bigr)^{-1}\!\int \! \! d\mathbf{x} \, \delta N^a \mathcal{D}_{a}^{\frak{g}(1)}, \label{eq:GeoDiffeoExpansionDef}
\end{equation}
where the perturbed density is
\begin{equation}
    \mathcal{D}_a^{\frak{g}(1)}=\gamma \biggl[\mathfrak{p}\biggl(\partial_a\delta K^b_b-\partial_i \delta K^i_a\biggr)-\mathfrak{c} \delta^j_a \partial_b \delta E^b_j\biggr].
\end{equation}

As for $\mathbb{G}$, the quantization of the diffeomorphism constraint in LQG via a group averaging procedure (see \cite{Han:2005km} for details) leads, in principle, to no quantum corrections — and, in particular, no GHCs — at the effective level.
However, recent results with corrected diffeomorphism constraints, such as a study of scalar cosmological perturbations in LQC using self-dual Ashtekar's variables \cite{BenAchour:2016leo} in which the usual signature change disappears, or in spherically symmetric models \cite{Arruga:2019kyd} where a no-go theorem is established on the \textit{non}-closure of the algebra of constraints, suggest that corrected diffeomorphism constraints are worth being considered, at least from a heuristic perspective. The study of corrected $\mathbb{D}_{\mathfrak{g}}$ is beyond the scope of this work and left for further research\footnote{The detailed calculations given in this work make this task easier. Specifically, most of the Poisson brackets presented in this study would remain unchanged under corrections of the diffeomorphism constraint. In addition, because of the definition of the Poisson brackets, part of $\left\{\mathbb{H}, \mathbb{D}\right\}$ can be derived performing the simple substitution $\frak{c}\rightarrow g(\frak{c}, \frak{p})$ for the anomalies at the perturbative level. This means that only the background brackets need to be calculated. Still, the calculation of $\left\{\mathbb{D}, \mathbb{D}\right\}$ and $\left\{\mathbb{D}, \mathbb{G}\right\}$ remains necessary and a specific study is required to ensure the closure of the algebra.}.

%% file: texts/constraints/geo_scalar.tex
 To ensure the full diffeomorphism invariance of the theory\footnote{In general, this statement holds true only on-shell, that is to say when the constraints are solved.}, another constraint, the so-called Hamiltonian (or scalar) constraint,  must be added  to the (spatial) diffeomorphism one. In particular, this constraint encodes the invariance of the theory under time reparametrization. When expressed in terms of Ashetekar's variables, modulo Gauss constraints, the gravitational part of the Hamiltonian constraint  reads \cite{Han:2005km}:
\begin{equation}
    \mathbb{H}_{\mathfrak{g}} \defeq \mathbb{H}_{\mathfrak{g}}[N] = \bigl( 2 \kappa \bigr)^{-1}\! \int \! \! d\mathbf{x} \, N \mathcal{H}_{\mathfrak{g}} = \bigl( 2 \kappa \bigr)^{-1}\! \int \! \! d\mathbf{x} \,  N \frac{E^c_j E^d_k}{\sqrt{\bigl| \text{det} E \bigr|}} \biggl[ \epsilon_i^{jk} F^i_{cd} - 2 \left( 1+ \gamma^2 \right) K^j_{[c}K^k_{d]}\biggr], \label{eq:HGeo}
\end{equation}
where $\mathcal{H}_{\mathfrak{g}}$ is the Hamiltonian (or scalar) constraint density and the field-strength $F^i_{ab}$ of the Ashtekar's connection $A^i_a$ is defined by:
\begin{equation}
    F^i_{ab} \defeq 2 \partial_{[a} A^i_{b]} + \epsilon^{i}_{jk} A^j_a A^k_b.
\end{equation}
The Hamiltonian density constraint perturbed at second order is decomposed as the sum of a background part,
\begin{equation}
    \mathcal{H}_\frak{g}^{(0)} = - 6 \sqrt{\frak{p}} \frak{c}^2,
\end{equation}
a first order term,
\begin{equation}
    \mathcal{H}_\frak{g}^{(1)} = - 4 \sqrt{\frak{p}} \delta K^b_b - \frac{\frak{c}^2}{\sqrt{\frak{p}}} \delta E^b_b + \frac{2}{\sqrt{\frak{p}}} \partial_a \partial^i \delta E^a_i,
\end{equation}
and a second order one,
\begin{align}
    \mathcal{H}_\frak{g}^{(2)} = &\sqrt{\frak{p}} \biggl[ \delta K^a_b \delta K^b_a - (\delta K^b_b) ^2 \biggr] - \frac12 \frac{\frak{c}^2}{ \frak{p}^{3/2}} \biggl[ \delta E^a_b \delta E^b_a - \frac12 (\delta E^b_b)^2 \biggr]  \notag \\
    &- 2 \frac{\frak{c}}{\sqrt{\frak{p}}} \delta K^i_a \delta E^a_i+ \frac{1}{\frak{p}^{3/2}} \mathcal{Z}_{ab}^{cidj} \bigl( \partial_c \delta E^a_i \bigr) \bigl( \partial_d \delta E^b_j \bigr).
\end{align}

In the previous expressions, the quantities
\begin{equation}
     \delta K^b_b \defeq \delta_i^a \delta K_a^i, \quad  \delta E^b_b \defeq \delta_a^i \delta E^a_i,
\end{equation}
and,
\begin{equation}
    \quad \delta K^a_b \delta K^b_a \defeq \delta^a_i \delta^b_j \delta K^j_a \delta K^i_b, \quad \delta E^a_b \delta E^b_a \defeq \delta^i_a \delta^j_b \delta E^a_j \delta E^b_i,
\end{equation}
have been introduced for simplicity and readiness. As in \cite{Cailleteau:2012fy}, we also define  $\mathcal{Z}_{ab}^{cidj}$ as:
\begin{equation}
     \mathcal{Z}_{ab}^{cidj} \defeq \frac14 \epsilon^{ef}_k \epsilon^k_{mn} \mathcal{X}^{mjd}_{be} \mathcal{X}^{nic}_{af} - \epsilon^{ie}_k \mathcal{X}^{kjd}_{bd} \delta^c_a - \epsilon^{ci}_k \mathcal{X}^{kjd}_{ba} + \frac12 \delta^i_a \epsilon^{ce}_k \mathcal{X}^{kjd}_{be},
\end{equation}
with,
\begin{equation}
    \mathcal{X}^{ijb}_{ca} \defeq \epsilon^{ij}_c \delta^b_a - \epsilon^{ib}_c \delta^j_a + \epsilon^{ijb} \delta_{ca} + \epsilon^{ib}_a \delta^j_c.
\end{equation}
It is argued in \cite{Cailleteau:2012fy} that the use of the scalar-vector-tensor decomposition of cosmological perturbations \cite{Lifshitz:1945du} would greatly simplify the expression of $\mathcal{Z}_{ab}^{cidj}$. Nonetheless, most conclusions can be reached without this step and, as far as the perturbation modes entering the calculation of the Poisson brackets are concerned, the present work remains generic. \\

The Hamiltonian constraint in the geometrical sector, perturbed at the second order, is therefore:
\begin{equation}
    \mathbb{H}_{\mathfrak{g}} = \bigl( 2 \kappa \bigr)^{-1}\! \int \! \! d\mathbf{x} \, \biggl( \mathbf{N} \biggl[\mathcal{H}_\frak{g}^{(0)}+\mathcal{H}_\frak{g}^{(2)}\biggr] + \delta N \mathcal{H}_\frak{g}^{(1)} \biggr).
\end{equation}
It should be noticed that, as long as background equations of motion are satisfied and for consistency of the perturbations theory, the first-order of the (geometrical) Hamiltonian constraint $\mathbb{H}_\frak{g}^{(1)}$ vanishes identically  \cite{Langlois:1994ec}. The explicit dependence of the previous constraint on the reduced curvature $\frak{c}$ at the perturbative level should be discussed as well. This becomes particularly important when constructing the effective theory using holonomy corrections. Historically, the correction has been applied to each occurrence of the reduced curvature $\frak{c}$ -- this has been the standard approach for discussing cosmological perturbation theory in the (effective) LQC context \cite{Mielczarek:2011ph, Cailleteau:2011kr, Cailleteau:2012fy, Linsefors:2012et, Cailleteau:2013kqa}. However, recent analysis have applied the corrections only to the background, $\mathcal{H}_\frak{g}^{(0)}$, leaving the perturbations $\mathcal{H}_\frak{g}^{(1)}$ and $\mathcal{H}_\frak{g}^{(2)}$ uncorrected ~\cite{Han:2017wmt}. This needs to be clarified as this might influence cosmological observables. In this work, we make explicit the reasons why applying the holonomy correction (either usual or generalized) to the perturbation expansion of the scalar constraint has no consequence at all when considering the perturbation theory at the second order.

%% file: texts/constraints/intro_scalar_field.tex
The coupling of the gravitational sector to scalar matter introduces additional contributions to the diffeomorphism and scalar constraints. 

%% file: texts/constraints/matter_diffeo.tex
Similarly to geometric phase-space variables, matter variables also transform under spatial diffeomorphisms. The invariance of the theory under those transformations is, in the canonical framework, ensured by the constraints. For a scalar field, the spatial diffeomorphism constraints are unique and expressed by:
\begin{equation}
    \mathbb{D}_{\mathfrak{m}} \defeq \mathbb{D}_{\mathfrak{m}}[N^a] = \int \! \! d\mathbf{x} \, N^a \mathcal{D}_{a}^{\mathfrak{m}} = \int \! \! d\mathbf{x} \, N^a \left( \pi_{\phi} \partial_a \phi \right). \label{eq:DmDef}
\end{equation}

At the background level the shift functions $\mathbf{N}^a$ vanish due to homogeneity and the perturbative expansion at second order of the matter diffeomorphism constraint is:
\begin{equation}
    \mathbb{D}_{\mathfrak{m}} = \int \! \! d\mathbf{x} \, \delta N^a \mathcal{D}_{a}^{\mathfrak{m}(1)}, \label{eq:DiffeoDensityMatterDef}
\end{equation}
where
\begin{equation}
    \mathcal{D}_{a}^{\mathfrak{m}(1)} = \boldsymbol{\pi} \partial_a \left( \delta \phi \right).
\end{equation}
This constraint does not depend on the reduced curvature $\mathfrak{c}$, so, independently of the way the diffeomorphism constraint is treated in LQG, no correction is applied to $\mathbb{D}_{\mathfrak{m}}$.

%% file: texts/constraints/matter_scalar.tex
The introduction of a scalar field leads to a modification of the Hamiltonian constraint. The matter contribution is given by \cite{Bojowald:2008gz}:
\begin{equation}
    \mathbb{H}_{\mathfrak{m}} \defeq \mathbb{H}_{\mathfrak{m}}[N] =\int \! \! d\mathbf{x} \, N \mathcal{H}_{\mathfrak{m}}=\int \! \! d\mathbf{x} \, N \left[ \frac{\pi_\phi^2}{2\sqrt{\bigl| \text{det} E \bigr|}}+\frac{E^a_i E^b_j }{2\sqrt{\bigl| \text{det} E \bigr|}}\delta^{ij}\partial_a \phi \partial_b \phi + \sqrt{\bigl| \text{det} E \bigr|} V(\phi)\right], \label{eq:HmDef}
\end{equation}
where $\mathcal{H}_\frak{m}$ corresponds to the matter Hamiltonian (or scalar) constraint density. Following~\cite{Bojowald:2008gz}, we write this density as:
\begin{equation}
    \mathcal{H}_{\mathfrak{m}} = \mathcal{H}_{\frak{m}, \pi}+\mathcal{H}_{\frak{m}, \nabla}+\mathcal{H}_{\frak{m}, \phi},
\end{equation}
where each term is respectively defined by:
\begin{equation}
    \mathcal{H}_{\frak{m}, \pi} \defeq \frac{\pi_\phi^2}{2\sqrt{\bigl| \text{det} E \bigr|}}, \quad \mathcal{H}_{\frak{m}, \nabla} \defeq \frac{E^a_i E^b_j }{2\sqrt{\bigl| \text{det} E \bigr|}}\delta^{ij}\partial_a \phi \partial_b \phi, \quad \text{ and } \quad \mathcal{H}_{\frak{m}, \phi} \defeq \sqrt{\bigl| \text{det} E \bigr|} V(\phi). \label{eq:ScalarMatterDensityDef}
\end{equation}
Is it important to stress that this constraint does not depend on any curvature-related quantity\footnote{It should be noticed that this statement would no longer hold if the counter-terms introduced in the next section to ensure the closure of the constraint algebra were functionally dependent on the curvature. As we will show, this specific scenario is not anomaly-free. However, this would be an implicit dependence, so no explicit correction should  be applied.}. The matter sector therefore does not require any holonomy correction. Determinants of the inverse densitized triad are however present and inverse-volume corrections are to be expected. More precisely,  $\mathcal{H}_{\frak{m}, \pi}$ and $\mathcal{H}_{\frak{m}, \nabla}$ should be inverse-volume corrected, but not $\mathcal{H}_{\frak{m}, \phi}$. We, however, do not consider such corrections in this work.
The derivation of consistency conditions with both holonomy and inverse-volume corrections has been adressed in \cite{Cailleteau:2013kqa} and the generalization of this work to GHCs is left for a future study.\\

The perturbative expansion of the matter Hamiltonian density $\mathcal{H}_\frak{m}$ can be readily obtained from Eq. (\ref{eq:ScalarMatterDensityDef}) \cite{Bojowald:2008gz}. Specifically, at the background level, the perturbative expansions are:
\begin{equation}
    \mathcal{H}_{\frak{m}, \pi}^{(0)} = \frac{\bm{\pi}^2}{2 \frak{p}^{3/2}}, \quad \mathcal{H}_{\frak{m}, \nabla}^{(0)} = 0, \quad \text{ and } \quad \mathcal{H}_{\frak{m}, \phi}^{(0)} = \frak{p}^{3/2} V[\bm{\phi}]. \label{eq:FirstOrderDensityMatterScal}
\end{equation}
The first order expansions read,
\begin{equation}
    \mathcal{H}_{\frak{m}, \pi}^{(1)} = \frac{\bm{\pi}}{\frak{p}^{3/2}} \, \delta \pi - \frac{\bm{\pi}^2}{4 \frak{p}^{5/2}} \, \delta E^b_b, \quad \mathcal{H}_{\frak{m}, \nabla}^{(1)} = 0, \quad \text{and} \quad \mathcal{H}_{\frak{m}, \phi}^{(1)} = \frak{p}^{3/2} \biggl( \partial_\phi \bigl( V[\bm{\phi}] \bigr) \delta \phi +  \frac{V[\bm{\phi}]}{2 \frak{p}} \delta E^b_b \biggr),
\end{equation}
and the second order expansion is,
\begin{equation}
    \mathcal{H}_{\frak{m}, \pi}^{(2)} = \frac{1}{2} \frak{p}^{-3/2} \biggl[ \delta \pi^2  -  \frac{\bm{\pi}}{\frak{p}} \delta \pi \delta E^b_b + \frac{\bm{\pi}^2}{4\frak{p}^2} \biggl(\frac12 (\delta E^b_b)^2 + \delta E^a_b \delta E^b_a \biggr) \biggr],
\end{equation}
\begin{equation}
     \mathcal{H}_{\frak{m}, \nabla}^{(2)} = \frac{\sqrt{\frak{p}}}{2} \delta^{ab} \, \partial_a (\delta \phi) \partial_b (\delta \phi),
\end{equation}
\begin{equation}
    \mathcal{H}_{\frak{m}, \phi}^{(2)}=\frak{p}^{3/2} \biggl[\frac12 \partial_\phi^2 \bigl( V[\bm{\phi}] \bigr) \delta \phi^2 +\frac{1}{2 \frak{p}} \partial_\phi \bigl( V[\bm{\phi}] \bigr) \delta \phi \, \delta E^b_b + \frac{V[\bm{\phi}]}{4 \frak{p}^2} \biggl( \frac12 (\delta E^b_b)^2 - \delta E^a_b \delta E^b_a\biggr) \biggr].
\end{equation}
 
The total matter contribution to the Hamiltonian constraint perturbed at second order is given by:
\begin{equation}
    \mathbb{H}_{\mathfrak{m}} = \bigl( 2 \kappa \bigr)^{-1}\! \int \! \! d\mathbf{x} \, \biggl( \mathbf{N} \biggl[\mathcal{H}_\frak{m}^{(0)}+\mathcal{H}_\frak{m}^{(2)}\biggr] + \delta N \mathcal{H}_\frak{m}^{(1)} \biggr).
\end{equation}

In perturbation theory, and as long as the background equations of motion are satisfied, the first-order expansion of the (matter) Hamiltonian $\mathbb{H}_\frak{m}^{(1)}$ vanishes \cite{Langlois:1994ec}, similarly to the gravitational sector. 

%% file: texts/constraints/algebra.tex
The algebra of constraints encodes the deep structure of the theory. 
In pure GR, it reads~\cite{Dirac:1958sc}:
\begin{align}
    \biggl\{\mathbb{D}[M^a], \mathbb{D}[N^b]\biggr\} &= \mathbb{D}\bigl[\mathcal{L}_{N^b} M^a\bigr],\\
    \biggl\{\mathbb{H}[M], \mathbb{D}[N^a]\biggr\} &= \mathbb{H}\bigl[\mathcal{L}_{N^a} M\bigr], \\
    \biggl\{\mathbb{H}[M], \mathbb{H}[N]\biggr\} &= \mathbb{D}\bigl[h^{ab}(M \nabla_b N - N \nabla_b M)  \bigr],
\end{align}  
where $h^{ab}$ is the (spatial) 3-metric, $\mathcal{L}_X f$ denotes the Lie derivative with respect to the $X$ vector field, $\mathbb{D}\defeq\mathbb{D}_\frak{g}+\mathbb{D}_\frak{m}$, and $\mathbb{H}\defeq\mathbb{H}_\frak{g}+\mathbb{H}_\frak{m}$. Holonomy-type corrections modify drastically the algebra of constraints and, consequently, the associated space-time structure~\cite{Teitelboim:1972vw}. Once effective (quantum) corrections have been applied, the Poisson brackets can be written as:
\begin{equation}
    \biggl\{\mathbb{C}^I, \mathbb{C}^J\biggr\} = f_K^{IJ} \, \mathbb{C}^K + \mathcal{A}^{\left\{ \mathbb{C}_I, \mathbb{C}_J \right\}},
\end{equation}
where $\mathcal{A}^{\left\{ \mathbb{C}_I, \mathbb{C}_J \right\}}$ are anomalous terms ``spoiling" the space-time structure and making the theory apparently inconsistent. To restore the consistency of the model, it is mandatory to close the algebra \cite{Teitelboim:1972vw}, i.e. to enforce $\mathcal{A}^{\left\{ \mathbb{C}^I, \mathbb{C}^J \right\}} = 0$. To this end, two distinct mathematical procedures have been so far considered. We study them in details in the following. \\

Recently, it has been elegantly suggested to take advantage of the \textit{a priori} freedom one has in the choice of the shape of the GHC to ensure the closure of the algebra \cite{Li:2023axl}. In practice, this involves deriving each $\mathcal{A}^{\left\{ \mathbb{C}^I, \mathbb{C}^J \right\}}$ and identifying a subset of GHCs that lead to $\mathcal{A}^{\left\{ \mathbb{C}^I, \mathbb{C}^J \right\}} = 0$. In the context of cosmological perturbation theory in LQC, it was shown in \cite{Li:2023axl} that the algebra can be closed for vector modes thanks to specific choices for the GHCs. This nice result is extended to scalar modes in Section (\ref{sec:closure_ghc}). We show that the procedure unfortunately fails and that another approach is required.\\

Historically introduced in \cite{Bojowald:2008gz} for LQC, and then considered extensively (see, e.g., \cite{Bojowald:2008jv, Wu:2012mh, Han:2018usc, BenAchour:2016leo, Mielczarek:2011ph, Cailleteau:2011kr, Cailleteau:2012fy, Cailleteau:2013kqa, Han:2017wmt}), another procedure to ensure the closure of the algebra of constraints involves the addition of counter-terms $\mathbb{C}^I_{ct}$ such that $\mathbb{C}^I \rightarrow \mathbb{C}^I + \mathbb{C}^I_{ct}$. One then has to find a peculiar $\mathbb{C}^I_{ct}$, \textit{i.e}, a peculiar deformation of the constraints, to ensure that $\mathcal{A}^{\left\{ \mathbb{C}^I, \mathbb{C}^J \right\}} = 0$. In this article, we parameterize the deformations following \cite{Cailleteau:2013kqa, Han:2017wmt}: 
\begin{equation}
    \mathbb{H}_{\mathfrak{g}/\mathfrak{m}}  \rightarrow \mathbb{H}_{\mathfrak{g}/\mathfrak{m}}  + \mathbb{H}_{\mathfrak{g}/\mathfrak{m}} ^{ct},
\end{equation}
where the additional terms are
\begin{equation}
\mathbb{H}^{ct}_{\mathfrak{g}/\mathfrak{m}} = \bigl( 2 \kappa \bigr)^{-1}\! \int \! \! d\mathbf{x} \, \biggl( \delta N \mathcal{H}_{\mathfrak{g}/\mathfrak{m}}^{(1)ct} + \mathbf{N} \mathcal{H}_{\mathfrak{g}/\mathfrak{m}}^{(2)ct} \biggr).
\end{equation}
In the previous expression, the densities of the gravitational sector are defined as
\begin{equation}
    \mathcal{H}_{\mathfrak{g}}^{(1)ct} = - 4 \alpha_1 \sqrt{\frak{p}} \delta K^b_b - \alpha_2\frac{\frak{c}^2}{\sqrt{\frak{p}}} \delta E^b_b + \frac{2}{\sqrt{\frak{p}}} \alpha_3 \partial_a \partial^i \delta E^a_i,
\end{equation}
at the first order and,
\begin{align}
    \mathcal{H}_\frak{g}^{(2)ct} = &\sqrt{\frak{p}} \biggl[ \alpha_4 (\delta K^a_b \delta K^b_a) - \alpha_5 (\delta K^b_b) ^2 \biggr] - \frac12 \frac{\frak{c}^2}{ \frak{p}^{3/2}} \biggl[\alpha_7 (\delta E^a_b \delta E^b_a) - \frac12 \alpha_8 (\delta E^b_b)^2 \biggr]  \notag \\
    &- 2 \frac{\frak{c}}{\sqrt{\frak{p}}} \alpha_6 (\delta K^i_a \delta E^a_i)+ \frac{\alpha_9}{\frak{p}^{3/2}} \mathcal{Z}_{ab}^{cidj} \bigl( \partial_c \delta E^a_i \bigr) \bigl( \partial_d \delta E^b_j \bigr),
\end{align}
at the second order.
The coefficients $\alpha_i\defeq\alpha_i(\frak{c}, \frak{p})$ appearing in the previous expressions are the so-called gravitational counter-terms. They are introduced to anticipate the appearance of anomalies that will, this way, be cancelled. They are required to have the correct classical limit and are assumed to always be factorized with terms already present in the original constraint. The variables upon which they depend is relevant since this will determine their contributions, or not, to the Poisson bracket defined on each variables of the phase-space. In \cite{Cailleteau:2011kr, Cailleteau:2013kqa},  and subsequent works, it was chosen to restrict the dependence of the counter-terms to the geometrical background phase-space variables only. We follow here this reasonable assumption. It is however important to underline that this restriction is mostly motivated by simplicity, without strong justification. This is another hypothesis that could be relaxed in future works. \\

In the matter sector, the deformation of the densities are given, at first order in perturbations, by
\begin{equation}
     \mathcal{H}_{\frak{m}}^{(1)ct} = \beta_1  \frac{\bm{\pi}}{\frak{p}^{3/2}} \delta \pi - \beta_2\frac{\bm{\pi}^2}{4 \frak{p}^{5/2}} \, \delta E^b_b + \frak{p}^{3/2} \biggl( \beta_3 \partial_\phi \bigl( V[\bm{\phi}] \bigr) \delta \phi + \beta_4 \frac{V[\bm{\phi}]}{2 \frak{p}} \delta E^b_b \biggr),
     \label{Hm with counter terms order 1}
\end{equation}
and, at the second order, by
\begin{align}
   \mathcal{H}_{\frak{m}}^{(2)ct} &= \frac{1}{2} \frak{p}^{-3/2} \biggl[\beta_5 \delta \pi^2  - \beta_6\frac{\bm{\pi}}{\frak{p}} \delta \pi \delta E^b_b + \frac{\bm{\pi}^2}{4\frak{p}^2} \biggl(\frac{\beta_7}{2} (\delta E^b_b)^2 + \beta_8 \delta E^a_b \delta E^b_a \biggr) \biggr] \notag \\
   &+ \beta_9 \frac{\sqrt{\frak{p}}}{2} \delta^{ab} \, \partial_a (\delta \phi) \partial_b (\delta \phi) + \frak{p}^{3/2} \biggl[\frac{\beta_{10}}{2} \partial_\phi^2 \bigl( V[\bm{\phi}] \bigr) \delta \phi^2 +\frac{\beta_{11}}{2 \frak{p}} \partial_\phi \bigl( V[\bm{\phi}] \bigr) \delta \phi \, \delta E^b_b \notag \\ 
   &+ \frac{V[\bm{\phi}]}{4 \frak{p}^2} \biggl( \frac{\beta_{12}}{2} (\delta E^b_b)^2 - \beta_{13}(\delta E^a_b \delta E^b_a)\biggr) \biggr]. 
   \label{Hm with counter terms order 2}
\end{align}
As for the gravitational sector, we have defined $\beta_i\defeq\beta_i(\frak{c}, \frak{p})$ as functions of the geometrical background variables only.

%% file: texts/brackets/G-G.tex
In the preceding sections, we have stated that the Gauss constraint undergoes no quantum corrections in LQG. Consequently, at the effective level, no anomaly is to be expected from the Poisson bracket $\left\{\mathbb{G}, \mathbb{G}\right\}$. Still, for the completeness of the generic aspect of this article we explicitly calculate this Poisson bracket step by step:
\begin{equation}
    \biggl\{\mathbb{G}, \mathbb{G}\biggr\}=\biggl\{\mathbb{G}, \mathbb{G}\biggr\}_{\mathfrak{c},\mathfrak{p}}+\biggl\{\mathbb{G}, \mathbb{G}\biggr\}_{\delta E,\delta K}+\biggl\{\mathbb{G}, \mathbb{G}\biggr\}_{\bm{\phi},\bm{\pi}}+\biggl\{\mathbb{G}, \mathbb{G}\biggr\}_{\delta \phi,\delta \pi}.
\end{equation}
Since the perturbed Gauss constraint does not depend on matter variables, see Eq. (\ref{eq:GaussExpansionDef}), the Poisson brackets from the matter sector vanish identically:
\begin{equation}
  \biggl\{\mathbb{G}, \mathbb{G}\biggr\}_{\bm{\phi},\bm{\pi}}=0 \qquad \text{and} \qquad \biggl\{\mathbb{G}, \mathbb{G}\biggr\}_{\delta \phi,\delta \pi}=0.
\end{equation}

Thus, only Poisson brackets from the geometrical sector need to be calculated. Considering the background part, it is clear from the construction of the second-order Gauss constraint Eq. (\ref{eq:GaussExpansionDef}) that derivatives of this constraint with respect to background variables would lead to second-order terms in the perturbative development. Therefore, the geometrical background part of the Poisson bracket is of order four in perturbations:
\begin{equation}
    \biggl\{\mathbb{G}, \mathbb{G}\biggr\}_{\mathfrak{c},\mathfrak{p}}=o(\delta \delta),
\end{equation}
where $\delta \delta$ denotes a product of \textit{first}-order perturbations. 
As the closure of the algebra is studied at the second order, those fourth-order terms are discarded. \\

Finally, the Poisson bracket of the geometrical sector at the level of the perturbed phase space must also be computed:
\begin{equation}
    \biggl\{\mathbb{G}[\Lambda^i], \mathbb{G}[\Gamma^j]\biggr\}_{\delta E,\delta K}=2\kappa^{-1}\!\int \!\! d \mathbf{x} \, \biggl[ \frak{cp} \biggl( \delta_{ij}  \delta \Lambda^i \delta \Gamma^j- \delta_{ij} \delta \Lambda^i \delta \Gamma^j \biggr) \biggr]=0.
\end{equation}
Therefore, as anticipated:
\begin{equation}
     \biggl\{\mathbb{G}, \mathbb{G}\biggr\}=0.
\end{equation}
Hence, no anomaly emerges from the bracket between two Gauss constraints.

%% file: texts/brackets/D-D.tex
Arguments similar to those established for the $\left\{ \mathbb{G}, \mathbb{G} \right\}$ bracket provide insights to the final outcome: aforementioned studies mentioned in Sec. (\ref{sec:geo_diffeo}) put aside, diffeomorphism constraints, both in the matter or geometric sectors, remain unaltered by holonomy corrections. Consequently, at this effective level, the classical result should be recovered, and no anomalies should arise from the Poisson bracket $\left\{ \mathbb{D}, \mathbb{D} \right\}$. Once again, for the sake of completeness, the computation is performed in details below.\\

To begin, it is interesting to notice that the geometric diffeomorphism constraint is independent of the matter sector, and the matter diffeomorphism constraint is independent of the geometric sector\footnote{The matter diffeomorphism constraint $\mathbb{D}_\frak{m}$ does depend on the perturbed shift functions $\delta N^a$, which is a geometric quantity. However, in the formulation considered in this work, the lapse and shift functions act as Lagrange multipliers on the constraints and are not phase space variables. An extended phase space including usual Lagrange multipliers can be considered, as discussed in \cite{Pons:1996av}.}. This means that, in full generality,
\begin{equation}
    \biggl\{ \mathbb{D}_\frak{g}, \mathbb{D}_\frak{m} \biggr\}=0.
\end{equation}
The brackets of the geometrical diffeomorphism constraints also vanish:
\begin{equation}
     \biggl\{ \mathbb{D}_\frak{g}, \mathbb{D}_\frak{g}\biggr\}_{\bm{\phi},\bm{\pi}}=0 \quad \text{ and } \quad \biggl\{ \mathbb{D}_\frak{g}, \mathbb{D}_\frak{g}\biggr\}_{\delta\phi, \delta\pi}=0,
\end{equation}
along with the brackets of the matter constraints:
\begin{equation}
    \biggl\{ \mathbb{D}_\frak{m}, \mathbb{D}_\frak{m}\biggr\}_{\frak{c}, \frak{p}}=0 \quad \text{ and } \quad \biggl\{ \mathbb{D}_\frak{m}, \mathbb{D}_\frak{m}\biggr\}_{\delta E, \delta K}=0.
\end{equation}
Moreover, in the matter sector, a comparison between the ADM and FLRW line elements presented Eqs. (\ref{eq:LineElemADM}) and (\ref{eq:LineElemFLRW}) shows that the shift functions $N^a$ vanish. Consequently, the perturbed matter diffeomorphism constraint is proportional to $\delta N^a$ only. The density is therefore perturbed \textit{only}  at the first-order  and is never proportional to either the full background canonical variables $\boldsymbol{\phi}$ and $\boldsymbol{\pi}$ or to the full perturbed canonical variables $\delta \phi$ and $\delta \pi$. It can therefore be concluded that
\begin{equation}
     \biggl\{ \mathbb{D}_\frak{m}, \mathbb{D}_\frak{m}\biggr\}_{\boldsymbol{\phi}, \boldsymbol{\pi}}=0 \quad \text{ and } \quad \biggl\{ \mathbb{D}_\frak{m}, \mathbb{D}_\frak{m}\biggr\}_{\delta\phi, \delta\pi}=0.
\end{equation}
From the perturbative construction of the (geometrical) diffeomorphism constraint at the second order, Eq. (\ref{eq:GeoDiffeoExpansionDef}), it can be seen that:
\begin{equation}
    \biggl\{ \mathbb{D}_\frak{g}, \mathbb{D}_\frak{g}\biggr\}_{\frak{c}, \frak{p}} = o(\delta \delta),
\end{equation}
leading to the same conclusion than previously. As for the perturbed geometrical aspect of the phase-space, the Poisson bracket is expressed as
\begin{eqnarray}
     \biggl\{ \mathbb{D}_\frak{g}[N_1^a], \mathbb{D}_\frak{g}[N_2^b]\biggr\}_{\delta E, \delta K} &=& \kappa^{-1}\!\int \!\! d \mathbf{x} \, \biggl[ \frak{cp} \biggl( \partial_a N_2^b \partial_b N_1^a + \partial_a N_2^a \partial_b N_1^b \\ \nonumber &&  - \partial_a N_2^b \partial_b N_1^a - \partial_a N_2^a \partial_b N_1^b\biggr) \biggr] \\ \nonumber 
     &=&0.
\end{eqnarray}
Thus, as anticipated,
\begin{equation}
     \biggl\{ \mathbb{D}, \mathbb{D}\biggr\}=0.
\end{equation}
No anomaly emerges from the bracket between two diffeomorphism constraints.

%% file: texts/brackets/D-G.tex
Since the Gauss constraint $\mathbb{G}$ depends neither on the scalar field $\phi$ nor on its momentum $\pi$, it can immediately be concluded, for the geometrical component of the diffeomorphism constraint, that
\begin{equation}
  \biggl\{\mathbb{D}_\frak{g}, \mathbb{G}\biggr\}_{\bm{\phi},\bm{\pi}}=0\quad \text{ and } \quad \biggl\{\mathbb{D}_\frak{g}, \mathbb{G}\biggr\}_{\delta \phi,\delta \pi}=0,
\end{equation}
and, similarly, in the matter counterpart, that,
\begin{equation}
  \biggl\{\mathbb{D}_\frak{m}, \mathbb{G}\biggr\}_{\bm{\phi},\bm{\pi}}=0\quad \text{ and } \quad \biggl\{\mathbb{D}_\frak{m}, \mathbb{G}\biggr\}_{\delta \phi,\delta \pi}=0.
\end{equation}
For the same reasons than for the $\left\{\mathbb{D}, \mathbb{D}\right\}$ bracket:
\begin{equation}
    \biggl\{ \mathbb{D}_\frak{m}, \mathbb{G}\biggr\}_{\frak{c}, \frak{p}}=0 \quad \text{ and } \quad \biggl\{ \mathbb{D}_\frak{m}, \mathbb{G}\biggr\}_{\delta E, \delta K}=0.
\end{equation}
And, as for $\left\{\mathbb{G}, \mathbb{G}\right\}$ and $\left\{\mathbb{D}, \mathbb{D}\right\}$, a careful examination of the second-order perturbative expansion of $\mathbb{D}_\frak{g}$ and $\mathbb{G}$ reveals that:
\begin{equation}
    \biggl\{\mathbb{D}_\frak{g}, \mathbb{G}\biggr\}_{\frak{c}, \frak{p}}=o(\delta \delta).
\end{equation}
 The last Poisson bracket to evaluate can be expressed as:
\begin{equation}
    \biggl\{\mathbb{D}_\frak{g}[N^a], \mathbb{G}[\Lambda^i]\biggr\}_{\delta E,\delta K}= \kappa^{-1} \int \!\! d \mathbf{x} \, \biggl[ - \frak{c}\frak{p} \biggl( \epsilon_{ib}^{\, \, a} \delta \Lambda^i\partial_a \delta N^b - \epsilon_{ib}^{\, \, a} \delta \Lambda^i\partial_a \delta N^b - \epsilon_{ji}^{\, \, \, i} \delta \Lambda^j \partial_a \delta N^a  \biggr) \biggr] =0,
\end{equation}
and is found to be identically zero. \\

In summary, as anticipated, the Poisson bracket $\left\{\mathbb{D}, \mathbb{G}\right\}$ yields
\begin{equation}
    \biggl\{\mathbb{D}, \mathbb{G}\biggr\} = 0.
\end{equation}
No anomaly arises here.

%% file: texts/brackets/H-G.tex
The most interesting brackets obviously involve the Hamiltonian constraint $\mathbb{H}$, which geometrical component $\mathbb{H}_\frak{g}$ receives, in principle, both holonomy and inverse-volume corrections while the matter part $\mathbb{H}_\frak{m}$ -- which does not depend upon the symmetry-reduced curvature $\frak{c}$ -- undergoes only inverse-volume modifications. This work still focuses only on the holonomy corrections.\\

As before, some elementary remarks allow to reduce the number of Poisson brackets to be calculated. In particular, the Gauss constraint is independent of the matter phase-space, both at the background and at the perturbations levels. Therefore, for the geometrical part of the constraint,
\begin{equation}
  \biggl\{\mathbb{H}_\frak{g}, \mathbb{G}\biggr\}_{\bm{\phi},\bm{\pi}}=0\quad \text{ and } \quad \biggl\{\mathbb{H}_\frak{g}, \mathbb{G}\biggr\}_{\delta \phi,\delta \pi}=0,
\end{equation}
and, for the matter component,
\begin{equation}
  \biggl\{\mathbb{H}_\frak{m}, \mathbb{G}\biggr\}_{\bm{\phi},\bm{\pi}}=0\quad \text{ and } \quad \biggl\{\mathbb{H}_\frak{m}, \mathbb{G}\biggr\}_{\delta \phi,\delta \pi}=0.
\end{equation}
Furthermore, derivatives of the Gauss constraint with respect to background quantities lead to terms perturbed at the second order. Hence, Poisson brackets related to the background phase space are reduced to Poisson brackets between the background Hamiltonian constraint and the Gauss constraint, that is
\begin{equation}
    \biggl\{\mathbb{H}, \mathbb{G} \biggr\}_{\frak{c}, \frak{p}} = \biggl\{\mathbb{H}^{(0)}, \mathbb{G} \biggr\}_{\frak{c}, \frak{p}} + o(\delta \delta).
\end{equation}
\subsubsection{Bracket $\left\{ \mathbb{H}_\frak{g}, \mathbb{G} \right\}$}
Thanks to the above simplifications, the Poisson bracket on the background part of the phase space can be computed quite straightforwardly:
\begin{subequations}
\begin{align}
    \biggl\{\mathbb{H}^{(0)}_\frak{g}[N], \mathbb{G}[\Lambda^i] \biggr\}_{\frak{c}, \frak{p}}&= \kappa^{-1}\!\int \!\! d \mathbf{x} \, \mathbf{N} \tensor{\epsilon}{_i_a^j} \delta \Lambda^i \delta E^a_j \biggl( \frac{g^2}{2 \sqrt{\frak{p}}} + 2 \sqrt{\frak{p}} g \partial_\frak{p} g\biggr) \\
    &+ \kappa^{-1}\!\int \!\! d \mathbf{x} \, \mathbf{N} \tensor{\epsilon}{_i_j^a} \delta \Lambda^i  \delta K^j_a \biggl( - 2\sqrt{\frak{p}}g \partial_\frak{c} g  \biggr).
\end{align}
\end{subequations}
This is clearly not related to a constraint. We shall come back later to the way this anomaly can be treated but we proceed, for now, with the calculation of the brackets. The contribution of the perturbed phase space to the full bracket is given by
\begin{subequations}
\begin{align}
    \biggl\{\mathbb{H}_\frak{g}[N], \mathbb{G}[\Lambda^i] \biggr\}_{\delta E, \delta K} &= \intwK  \mathbf{N} \, \tensor{\epsilon}{_a_i^j} \delta \Lambda^i \delta E^a_j \biggl( \frac{\frak{c}}{\sqrt{\frak{p}}} \bigl[ \tilde{g} + \alpha_6 \bigr] - \frac{1}{\sqrt{\frak{p}}}\bigl[\tilde{g}^2 + \alpha_7\bigr] \biggr) \\
    &+ \intwK \mathbf{N} \, \tensor{\epsilon}{_j_i^a} \delta \Lambda^i  \delta K^j_a \biggl( -\frak{c} \sqrt{\frak{p}} \bigl[ 1 + \alpha_4 \bigr] -  \sqrt{\frak{p}} \bigl[ \tilde{g} + \alpha_6 \bigr] \biggr),
\end{align}
\end{subequations}
which completes the computation of $\left\{ \mathbb{H}_\frak{g}, \mathbb{G}\right\}$. Gathering everything together leads to:
\begin{equation}
     \biggl\{\mathbb{H}_\frak{g}[N], \mathbb{G}[\Lambda^i]\biggr\} = \kappa^{-1}\!\int \!\! d \mathbf{x} \, \frac{N}{2 \sqrt{p}} \epsilon_{ia}^{\, \, \,j} \delta E^a_j \delta \Lambda^i \mathcal{A}_1^{\left\{ \mathbb{H}, \mathbb{G} \right\}} + \kappa^{-1}\!\int \!\! d \mathbf{x} \, N \sqrt{p} \epsilon_{ij}^{\, \, \,a} \delta K^j_a \delta \Lambda^i \mathcal{A}_2^{\left\{ \mathbb{H}, \mathbb{G} \right\}} \label{eq:HgGFull},
\end{equation}
where the anomalies are given by
\begin{equation}
     \mathcal{A}_1^{\left\{ \mathbb{H}, \mathbb{G} \right\}}= g^2 + 2 \frak{c} \bigl[ \tilde{g} + \alpha_6 \bigr] - \bigl[\tilde{g}^2 + \alpha_7\bigr]+ 4 \frak{p} g \partial_\frak{p} g, \label{eq:AHG1}
\end{equation}
and
\begin{equation}
    \mathcal{A}_2^{\left\{ \mathbb{H}, \mathbb{G} \right\}}=\frak{c}\bigl[ 1 + \alpha_4 \bigr]+\bigl[\tilde{g} + \alpha_6\bigr] - 2 g \partial_\frak{c} g. \label{eq:AHG2}
\end{equation}
\subsubsection{Bracket $\left\{ \mathbb{H}_\frak{m}, \mathbb{G} \right\}$}

Due to the presence of counter-terms in $\mathbb{H}_\frak{m}$, see Eqs. (\ref{Hm with counter terms order 1}) and (\ref{Hm with counter terms order 2}), the geometrical parts of the bracket $\left\{ \mathbb{H}_\frak{m}, \mathbb{G} \right\}$ might not vanish and need to be calculated, even though the constraint undergoes no holonomy corrections. \\

The bracket for the geometrical background phase space canonical variables reads:
\begin{align}
    \biggl\{\mathbb{H}^{(0)}_\frak{m}[N], \mathbb{G}[\Lambda] \biggr\}_{\frak{c}, \frak{p}} &= \intnK \mathbf{N} \, \tensor{\epsilon}{_i_a^j} \delta \Lambda^i \delta E^a_j \biggl( \frac{\bm{\pi}^2}{4 \frak{p}^{5/2}} - \frac{\sqrt{\frak{p}}}{2} V(\bm{\phi}) \biggr),
\end{align}
while it is, at the geometrical perturbed level, 
\begin{subequations}
\begin{align}
     \biggl\{\mathbb{H}_\frak{m}[N], \mathbb{G}[\Lambda^i] \biggr\}_{\delta E, \delta K} &= \intnK \mathbf{N} \, \tensor{\epsilon}{_i_a^j} \delta \Lambda^i \delta E^a_j \biggl( -\frac{\bm{\pi}^2}{4 \frak{p}^{5/2}} + \frac{\sqrt{\frak{p}}}{2} V(\bm{\phi}) \biggr) \label{eq:HmGFull1}\\
     &+ \intnK \frac{\mathbf{N} \bm{\pi}^2}{4 \frak{p}^{5/2}} \, \tensor{\epsilon}{_i_a^j} \delta \Lambda^i \delta E^a_j  \mathcal{A}_3^{\left\{ \mathbb{H}, \mathbb{G} \right\}} \\
     &+ \intnK \frac{\mathbf{N} \sqrt{\frak{p}}}{2} V(\bm{\phi}) \, \tensor{\epsilon}{_i_a^j} \delta \Lambda^i \delta E^a_j  \mathcal{A}_4^{\left\{ \mathbb{H}, \mathbb{G} \right\}} \label{eq:HmGFull2}.
\end{align}
\end{subequations}
The two new anomalies are:
\begin{equation}
    \mathcal{A}_3^{\left\{ \mathbb{H}, \mathbb{G} \right\}}=-\beta_{10} \quad \text{ and } \quad \mathcal{A}_4^{\left\{ \mathbb{H}, \mathbb{G} \right\}} = \beta_{12}. \label{eq:AHG34}
\end{equation}
Hence, the bracket $\left\{ \mathbb{H}, \mathbb{G}\right\}$ leads to four new anomalies $\mathcal{A}_1^{\left\{ \mathbb{H}, \mathbb{G} \right\}},~...~,\mathcal{A}_4^{\left\{ \mathbb{H}, \mathbb{G} \right\}}$.

%% file: texts/brackets/H-D.tex
Thanks to the linearity of the Poisson brackets and using the definitions of the total Hamiltonian and diffeomorphism constraints, $\mathbb{H}$ and  $\mathbb{D}$, the full $\left\{ \mathbb{H}, \mathbb{D}\right\}$ bracket can be splitted as:
\begin{equation}
    \biggl\{ \mathbb{H}, \mathbb{D}\biggr\} =\biggl\{ \mathbb{H}_\frak{g}, \mathbb{D}_\frak{g}\biggr\} +\biggl\{ \mathbb{H}_\frak{m}, \mathbb{D}_\frak{m}\biggr\} +\biggl\{ \mathbb{H}_\frak{g}, \mathbb{D}_\frak{m}\biggr\} +\biggl\{ \mathbb{H}_\frak{m}, \mathbb{D}_\frak{g}\biggr\} .
\end{equation}
This allows some straightforward simplifications. In particular, Eqs. (\ref{eq:HGeo}) and (\ref{eq:DmDef}) show that the geometrical part of the Hamiltonian constraint, that is $\mathbb{H}_\frak{g}$, depends only on the geometrical phase space whereas the matter part of the diffeomorphism constraint, that is $\mathbb{D}_\frak{m}$, depends only on the variables related to the scalar field. Hence,
\begin{equation}
    \biggl\{ \mathbb{H}_\frak{g}, \mathbb{D}_\frak{m}\biggr\}  = 0,
\end{equation}
and
\begin{equation}
    \biggl\{ \mathbb{H}_\frak{m}, \mathbb{D}_\frak{m}\biggr\} _{\frak{c}, \frak{p}} = 0, \quad \text{ together with } \quad \biggl\{ \mathbb{H}_\frak{m}, \mathbb{D}_\frak{m}\biggr\} _{\delta E, \delta K} = 0.
\end{equation}

In addition, as $\mathbb{D}_\frak{g}$ neither depends on the canonical variables of the scalar field, one obtains, for the geometrical sector,
\begin{equation}
    \biggl\{ \mathbb{H}_\frak{g}, \mathbb{D}_\frak{g}\biggr\} _{\bm{\phi}, \bm{\pi}} = 0, \quad \text{ and } \quad \biggl\{ \mathbb{H}_\frak{g}, \mathbb{D}_\frak{g}\biggr\} _{\delta \phi,\delta \pi} = 0
\end{equation}
and, for the matter sector,
\begin{equation}
    \biggl\{ \mathbb{H}_\frak{m}, \mathbb{D}_\frak{g}\biggr\} _{\bm{\phi}, \bm{\pi}} = 0 \quad \text{ and } \quad \biggl\{ \mathbb{H}_\frak{m}, \mathbb{D}_\frak{g}\biggr\} _{\delta \phi,\delta \pi} = 0.
\end{equation}
As a final simplification at this stage, it can also be remarked that 
\begin{equation}
    \biggl\{\mathbb{H}, \mathbb{D} \biggr\}_{\frak{c}, \frak{p}} = \biggl\{\mathbb{H}^{(0)}, \mathbb{D} \biggr\}_{\frak{c}, \frak{p}} + o(\delta \delta).
\end{equation}
\subsubsection{Bracket $\left\{ \mathbb{H}_\frak{g}, \mathbb{D}_\frak{g} \right\}$} 
A simple computation yields:
\begin{subequations}
\begin{align}
    \biggl\{\mathbb{H}^{(0)}_\frak{g}[N], \mathbb{D}_\frak{g}[N^a] \biggr\}_{\frak{c}, \frak{p}} &= \intwK \mathbf{N} \, \delta N^a \partial_b \left(  \delta^i_a \delta E^b_i \right) \biggl( - \frac{g^2}{2 \sqrt{\frak{p}}} - 2 \sqrt{\frak{p}} g \partial_\frak{p} g\biggr) \\
    &+ \intwK \mathbf{N} \, \delta N^a \partial_b \left(  \delta^b_i \delta K^i_a \right) \biggl( 2 \sqrt{\frak{p}} g \partial_\frak{c}g \biggr) \\
    &+  \intwK \mathbf{N} \, \delta N^a \partial_a \left( \delta K^b_b \right) \biggl( - 2 \sqrt{\frak{p}} g  \partial_\frak{c}g \biggr),
\end{align}
\end{subequations}
and, for the geometrical perturbations,
\begin{subequations}
\begin{align}
    \biggl\{\mathbb{H}_\frak{g}[N], \mathbb{D}_\frak{g}[N^a] \biggr\}_{\delta E, \delta K} &= \intwK \delta N \partial_a \delta N^a \biggl(  - \sqrt{\frak{p}}\bigl[\tilde{g}^2 + \alpha_2 \bigr] - 2 \sqrt{\frak{p}}\frak{c} \bigl[ \tilde{g} + \alpha_1 \bigr]\biggr) \\
    &+ \intwK \mathbf{N} \sqrt{\frak{p}} \, \partial_a \bigl(\delta N^b\bigr) \delta^a_i \delta K^i_b \biggl( \frak{c}\bigl[1 + \alpha_4 \bigr] + \bigl[ \tilde{g} + \alpha_6 \bigr] \biggr) \\
    &+ \intwK \mathbf{N} \sqrt{\frak{p}} \, \partial_a \bigl(\delta N^a\bigr)  \delta K^b_b \biggl( -\frak{c}\bigl[1 + \alpha_5 \bigr] + \bigl[ \tilde{g} + \alpha_6 \bigr]\biggr) \\
    &+ \intwK \frac{\mathbf{N}}{2 \sqrt{\frak{p}}} \, \partial_a \bigl( \delta N^b \bigr) \delta^i_b \delta E^a_i \biggl( \bigl[ \tilde{g}^2 + \alpha_7\bigr] - 2 \frak{c} \bigl[ \tilde{g} + \alpha_6 \bigr] \biggr)\\
    &+ \intwK \frac{\mathbf{N}}{2 \sqrt{\frak{p}}} \, \partial_a \bigl( \delta N^a \bigr) \delta E^b_b \biggl( \alpha_8 - \alpha_7 \biggr).
\end{align}
\end{subequations}
It is important to notice that, in the above reduced brackets of the geometrical components of the constraints, the (corrected) geometrical Hamiltonian constraint is not recovered. In other words, because of the holonomy correction, the somehow expected structure is not ensured at the perturbative level. However, if loop corrections are applied \textit{only} at the background level (as in \cite{Han:2017wmt}), the classical result is recovered.\\

When summing up everything and after integrating by parts, one is led to:
\begin{subequations}
\begin{align}
    \biggl\{\mathbb{H}_\frak{g}[N], \mathbb{D}_\frak{g}[N^a] \biggr\} &= \mathbb{H}_\frak{g}[\delta N \partial_a \delta N^a] \label{eq:HDFull1}\\ 
     &+ \intwK \sqrt{\frak{p}} \, \delta N \partial_a \bigl( \delta N^a \bigr) \mathcal{A}_1^{{\left\{ \mathbb{H}, \mathbb{D} \right\}}}\\
     &+ \intwK \frac{\mathbf{N}}{2\sqrt{\frak{p}}} \,\partial_a \bigl(\delta N^a\bigr) \delta E^b_b \mathcal{A}_2^{{\left\{ \mathbb{H}, \mathbb{D} \right\}}} \\
     &+ \intwK \mathbf{N} \sqrt{\frak{p}} \,\partial_a \bigl(\delta N^a\bigr) \delta K^b_b \mathcal{A}_3^{{\left\{ \mathbb{H}, \mathbb{D} \right\}}} \\
     &+ \intwK \frac{\mathbf{N}}{2\sqrt{\frak{p}}} \,\partial_a \bigl(\delta N^b\bigr) \delta^i_b \delta E^a_i \mathcal{A}_4^{{\left\{ \mathbb{H}, \mathbb{D} \right\}}} \\
     &+ \intwK \mathbf{N} \sqrt{\frak{p}} \,\partial_a \bigl(\delta N^b\bigr) \delta^a_i \delta K^i_b  \mathcal{A}_5^{{\left\{ \mathbb{H}, \mathbb{D} \right\}}} \label{eq:HDFull2},
\end{align}
\label{eq:HDFull}
\end{subequations}
where the anomalies are: 
\begin{align}
    \mathcal{A}_1^{{\left\{ \mathbb{H}, \mathbb{D} \right\}}} &= 3 g^2 -  \bigl[\tilde{g}^2 + \alpha_2 \bigr] - 2 \frak{c} \bigl[ \tilde{g} + \alpha_1  \bigr], \label{eq:AHD1} \\
    \mathcal{A}_2^{{\left\{ \mathbb{H}, \mathbb{D} \right\}}}&= \alpha_8 - \alpha_7,\label{eq:AHD2}\\
    \mathcal{A}_3^{{\left\{ \mathbb{H}, \mathbb{D} \right\}}} &= \frak{c}\bigl[1 + \alpha_5 \bigr] + \bigl[\tilde{g} + \alpha_6 \bigr] - 2 g \partial_\frak{c}g, \label{eq:AHD3} \\
    \mathcal{A}_4^{{\left\{ \mathbb{H}, \mathbb{D} \right\}}} &= g^2 - 2 \frak{c}\bigl[ \tilde{g} + \alpha_6 \bigr] + \bigl[\tilde{g}^2 + \alpha_7 \bigr] + 4 \frak{p} g \partial_\frak{p}g, \label{eq:AHD4}  \\
    \mathcal{A}_5^{{\left\{ \mathbb{H}, \mathbb{D} \right\}}} &= \frak{c}\bigl[1 + \alpha_4] + \bigl[\tilde{g}  + \alpha_6\bigr] - 2 g \partial_\frak{c}g.
    \label{eq:AHD5}
\end{align}

It can immediately be noticed that the anomalies $\mathcal{A}_1^{{\left\{ \mathbb{H}, \mathbb{G} \right\}}}$ (\ref{eq:AHG1}) and $ \mathcal{A}_2^{{\left\{ \mathbb{H}, \mathbb{G} \right\}}}$ (\ref{eq:AHG2}) coming from the $\left\{\mathbb{H}_\frak{g}, \mathbb{G}\right\}$ bracket are, respectively, equal to $\mathcal{A}_4^{{\left\{ \mathbb{H}, \mathbb{D} \right\}}}$ (\ref{eq:AHD4}) and $\mathcal{A}_5^{{\left\{ \mathbb{H}, \mathbb{G} \right\}}}$ (\ref{eq:AHD5}). In other words, the bracket $\left\{\mathbb{H}_\frak{g}, \mathbb{G}\right\}$ is redundant with $\left\lbrace\mathbb{H}_\frak{g}, \mathbb{D}_\frak{g} \right\rbrace$ from the viewpoint of anomaly cancellation. This is due, on the one hand, to the structural similarities of the geometrical diffeomorphism constraint $\mathbb{D}_\frak{g}$ and Gauss constraint $\mathbb{G}$ at the perturbative level, and, on the other hand, to the fact that none of those constraints receives loop corrections. Should the diffeomorphism constraint be corrected, this conclusion might not hold anymore.  

\subsubsection{Bracket $\left\{ \mathbb{H}_\frak{m}, \mathbb{D}_\frak{m} \right\}$}
As discussed previously, only Poisson brackets of the scalar field part of the phase space need to be computed to get the full result for $\left\{ \mathbb{H}_\frak{m}, \mathbb{D}_\frak{m} \right\}$. This greatly simplifies the analysis. 
Let us start with the background pair of canonical variables $(\bm{\phi}, \bm{\pi})$:
\begin{align}
    \biggl\{\mathbb{H}^{(0)}_\frak{m}[N], \mathbb{D}_\frak{m}[N^a] \biggr\}_{\bm{\phi}, \bm{\pi}} &= \intnK \mathbf{N} \, \frak{p}^{3/2} \partial_\phi \bigl( V[\bm{\phi}]  \bigr)\, \delta N^a \partial_a \bigl( \delta \phi \bigr).
\end{align}
At the level of the perturbed phase space, one gets:
\begin{subequations}
\begin{align}
    \biggl\{\mathbb{H}_\frak{m}[N], \mathbb{D}_\frak{m}[N^a] \biggr\}_{\delta \phi, \delta \pi} &= \intnK \frac{\bm{\pi}^2}{\frak{p}^{3/2}} \, \delta N \partial_a \bigl( \delta N^a \bigr) \biggl( 1+\beta_1\biggr) \label{eq:Kin1} \\
    &+ \intnK \frac{\mathbf{N} \bm{\pi}}{\frak{p}^{3/2}} \, \partial_a \bigl( \delta N^a \bigr)  \delta \pi \biggl(  1+\beta_5 \biggr) \\
    &+ \intnK \frac{\mathbf{N} \bm{\pi}^2}{2\frak{p}^{5/2}} \, \partial_a \bigl( \delta N^a \bigr) \delta E^b_b \biggl( -1 - \beta_6 \biggr).
\end{align}
\end{subequations}
At this point, no obvious conclusion can be reached as it not clear whether this term is anomalous or not.  In addition, contrary to what could have been naively expected, the scalar matter constraint is not fully recovered from this bracket. However, the situation is different from the one of the bracket $\left\{ \mathbb{H}_\frak{g}, \mathbb{D}_\frak{g} \right\}$: no specific procedure is here necessary to exhibit the scalar matter constraint as it naturally appears as an interplay between the geometrical diffeomorphism constraint and the scalar matter constraint in the computation of $\left\{ \mathbb{H}_\frak{m}, \mathbb{D}_\frak{g} \right\}$.

\subsubsection{Bracket $\left\{ \mathbb{H}_\frak{m}, \mathbb{D}_\frak{g} \right\}$}
This is the last step in the calculation of the full  $\left\{ \mathbb{H}, \mathbb{D} \right\}$ bracket. As explained earlier, it is not necessary to compute the Poisson brackets on the matter phase space.  Still, the brackets on the geometrical part of the phase space have to be calculated. Starting with the background pair $\bigl( \frak{c}, \frak{p} \bigr)$ of canonical variables:
\begin{align}
    \biggl\{\mathbb{H}^{(0)}_\frak{m}[N], \mathbb{D}_\frak{g}[N^a] \biggr\}_{\frak{c}, \frak{p}} &= - \intnK \frac{\mathbf{N} \bm{\pi}^2}{4 \frak{p}^{5/2}} \, \delta N^a \partial_b \bigl( \delta^i_a \delta E^b_i \bigr) \\
    &+ \intnK \frac{\mathbf{N} \sqrt{\frak{p}}}{2} V[\bm{\phi}] \, \delta N^a \partial_b \bigl( \delta^i_a \delta E^b_i \bigr),
\end{align}
while, at the perturbed level, one gets
\begin{subequations}
\begin{align}
    \biggl\{\mathbb{H}_\frak{m}[N], \mathbb{D}_\frak{g}[N^a] \biggr\}_{\delta E, \delta K} &= \intnK \frac{\bm{\pi}^2}{2 \frak{p}^{3/2}} \, \delta N \partial_a \bigl( \delta N^a \bigr) \biggl( -1 - \beta_2 \biggr) \label{eq:Kin2} \\
    &+ \intnK \frac{\mathbf{N} \bm{\pi}}{\frak{p}^{3/2}} \, \partial_a \bigl( \delta N^a \bigr) \delta \pi \biggl( -1 - \beta_6 \biggr) \\
    &+ \intnK \frac{\mathbf{N} \bm{\pi}^2}{4 \frak{p}^{5/2}}  \partial_a \bigl( \delta N^b \bigr) \delta^i_b \delta E^a_i \biggl( -1 - \beta_{10} \biggr) \\
    &+ \intnK \frac{\mathbf{N} \sqrt{\frak{p}}}{2} V[\bm{\phi}] \, \partial_a \bigl( \delta N^b \bigr) \delta^i_b \delta E^a_i \biggl( 1 + \beta_{12} \biggr) \\
    &+ \intnK \frac{\mathbf{N} \bm{\pi}^2}{4 \frak{p}^{5/2}} \, \partial_a \bigl( \delta N^a \bigr) \delta E^b_b \biggl( 2 + \beta_{10} + \beta_9 \biggr) \\
    &+ \intnK \frac{\mathbf{N} \sqrt{\frak{p}}}{2} V[\bm{\phi}]  \, \partial_a \bigl( \delta N^a \bigr) \delta E^b_b \biggl( \beta_{11} - \beta_{12} \biggr) \\
    &+ \intnK \mathbf{N} \, \frak{p}^{3/2} \partial_\phi \bigl( V[\bm{\phi}] \bigr) \, \partial_a \bigl(  \delta N^a \bigr) \delta \phi \biggl( 1 + \beta_8 \biggr) \\
    &+ \intnK \frak{p}^{3/2}\,  V[\bm{\phi}] \, \delta N  \partial_a \bigl(  \delta N^a \bigr) \biggl( 1 + \beta_4 \biggr). \label{eq:Pot1}
\end{align}
\end{subequations}
It should be noticed that, combining Eqs. (\ref{eq:Kin1}) and (\ref{eq:Kin2}), the kinetic term of the matter scalar constraint, Eq. (\ref{eq:FirstOrderDensityMatterScal}), is obtained, modulo an anomaly, while the potential part of the constraint is given, modulo $\beta_4$, by Eq. (\ref{eq:Pot1}). In other words, the expected classical result for the matter sector is recovered via the bracket $\left\{ \mathbb{H}_\frak{m}, \mathbb{D}_\frak{g} + \mathbb{D}_\frak{m}\right\}$. This is a key difference with the geometrical sector where a specific procedure was necessary. \\ 

It can now be concluded:
\begin{subequations}
\begin{align}
    \biggl\{\mathbb{H}_\frak{m}[N], \mathbb{D}_\frak{g}[N^a]+\mathbb{D}_\frak{m}[N^a]\biggr\} &= \mathbb{H}_\frak{m}[\delta N \partial_a \delta N^a] \label{eq:HDtot1}\\
    &+ \intnK \frac{\mathbf{N} \bm{\pi}}{\frak{p}^{3/2}} \, \partial_a \bigl( \delta N^a \bigr) \delta \pi  \mathcal{A}_6^{{\left\{ \mathbb{H}, \mathbb{D} \right\}}} \\
     &+ \intnK \frac{\bm{\pi}^2}{2 \frak{p}^{3/2}} \, \delta N \partial_a \bigl( \delta N^a \bigr)\mathcal{A}_7^{{\left\{ \mathbb{H}, \mathbb{D} \right\}}} \\
     &+ \intnK \frak{p}^{3/2}\, V[\bm{\phi}] \, \delta N \partial_a \bigl( \delta N^a \bigr) \mathcal{A}_8^{{\left\{ \mathbb{H}, \mathbb{D} \right\}}} \\
     &+ \intnK \frac{\mathbf{N} \bm{\pi}^2}{4 \frak{p}^{5/2}} \, \partial_a \bigl( \delta N^b \bigr) \delta^i_b \delta E^a_i \mathcal{A}_9^{{\left\{ \mathbb{H}, \mathbb{D} \right\}}} \\
     &+ \intnK \frac{\mathbf{N} \sqrt{\frak{p}}}{2} V[\bm{\phi}] \, \partial_a \bigl( \delta N^b \bigr) \delta^i_b \delta E^a_i \mathcal{A}_{10}^{{\left\{ \mathbb{H}, \mathbb{D} \right\}}} \\
     &+ \intnK \frac{\mathbf{N} \bm{\pi}}{4 \frak{p}^{5/2}} \, \partial_a \bigl( \delta N^a \bigr) \delta E^b_b \mathcal{A}_{11}^{{\left\{ \mathbb{H}, \mathbb{D} \right\}}}\\
     &+ \intnK \frac{\mathbf{N} \sqrt{\frak{p}}}{2} \, \partial_\phi \bigl( V[\bm{\phi}] \bigr) \,  \partial_a \bigl( \delta N^a \bigr)\delta \phi \mathcal{A}_{12}^{{\left\{ \mathbb{H}, \mathbb{D} \right\}}}\\
     &+  \intnK \frac{\mathbf{N} \sqrt{\frak{p}}}{2} \, V[\bm{\phi}] \partial_a \bigl( \delta N^a \bigr) \delta E^b_b \mathcal{A}_{13}^{{\left\{ \mathbb{H}, \mathbb{D} \right\}}}. \label{eq:HDtot2}
\end{align}
\label{eq:HDtot}
\end{subequations}
The anomalies are defined by:
\begin{align}
     \mathcal{A}_6^{{\left\{ \mathbb{H}, \mathbb{D} \right\}}} &= \beta_5 - \beta_6, \label{eq:AHD6}\\
     \mathcal{A}_7^{{\left\{ \mathbb{H}, \mathbb{D} \right\}}} &= 2 \beta_1 - \beta_2, \label{eq:AHD7}\\
      \mathcal{A}_8^{{\left\{ \mathbb{H}, \mathbb{D} \right\}}} &= \beta_4, \label{eq:AHD8}\\
      \mathcal{A}_9^{{\left\{ \mathbb{H}, \mathbb{D} \right\}}} &= - \beta_{10}, \label{eq:AHD9}\\
      \mathcal{A}_{10}^{{\left\{ \mathbb{H}, \mathbb{D} \right\}}} &= \beta_{12}, \label{eq:AHD10}\\
        \mathcal{A}_{11}^{{\left\{ \mathbb{H}, \mathbb{D} \right\}}} &= \beta_{10} - 2 \beta_6 + \beta_9, \label{eq:AHD11}\\
        \mathcal{A}_{12}^{{\left\{ \mathbb{H}, \mathbb{D} \right\}}} &= \beta_8, \label{eq:AHD12}\\
        \mathcal{A}_{13}^{{\left\{ \mathbb{H}, \mathbb{D} \right\}}} &= \beta_{11} - \beta_{12}. \label{eq:AHD13}
\end{align}
As for the geometrical part $\mathbb{H}_\frak{g}$ of the Hamiltonian constraint $\mathbb{H}$, it should be noticed that $\mathcal{A}_3^{{\left\{ \mathbb{H}, \mathbb{G} \right\}}}$ (\ref{eq:AHG34}) and $\mathcal{A}_4^{{\left\{ \mathbb{H}, \mathbb{G} \right\}}}$ (\ref{eq:AHG34}) are, respectively, equal to $\mathcal{A}_9^{{\left\{ \mathbb{H}, \mathbb{D} \right\}}}$ (\ref{eq:AHD9}) and $\mathcal{A}_{10}^{{\left\{ \mathbb{H}, \mathbb{D} \right\}}}$ (\ref{eq:AHD10}). Hence, there are only six new anomalies, leading to a total amount of nine anomalies arising from the bracket $\left\{ \mathbb{H}, \mathbb{D}\right\}$. It is interesting to underline that, until now, all the anomalies  are either depending on the counter-terms from the geometrical sector or on the counter-terms from the matter sector. This obviously simplifies the closure of the algebra  (sometimes trivially by setting counter-terms to zero, which means that the bracket would have naturally been anomaly-free). This property is however lost when considering the anomalies appearing in the bracket $\left\{ \mathbb{H}, \mathbb{H}\right\}$, as we will see right now.

%% file: texts/brackets/H-H.tex
\newcommand{\ddN}{\Delta ( \delta N )}
\newcommand{\pc}[1]{\partial_\frak{c} \bigl( #1 \bigr)}
\newcommand{\pp}[1]{\partial_\frak{p} \bigl( #1 \bigr)}
\newcommand{\ppc}[1]{\partial_\frak{c} #1 }
\newcommand{\ppp}[1]{\partial_\frak{p} #1 }
Let us now proceed with the calculation of the most interesting (and computationally intricate) bracket. As we shall see, all  the previously considered brackets ultimately reduce to the algebra of constraints of GR. However, the $\left\{ \mathbb{H}, \mathbb{H} \right\}$ bracket introduces a fundamental deformation of the algebra, which could significantly change cosmological predictions \cite{Cailleteau:2011kr, Linsefors:2012et, Bolliet:2015bka, Renevey:2021tmh, DeSousa:2022rep}. Using  the linearity of the  brackets and the definition of the constraints, one straightforwardly gets\footnote{The calculation of the bracket $\left\{ \mathbb{H}, \mathbb{H} \right\}$ requires to smear the scalar constraints with two different lapse function $N_1$ and $N_2$. In particular, $\delta N_1$ and $\delta N_2$ have no reason to be the same. Still, for the background, a single lapse $\mathbf{N}$ must be used.}:
\begin{align}
    \biggl\{\mathbb{H}[N_1], \mathbb{H}[N_2]\biggr\} = &\biggl\{\mathbb{H}_\frak{g}[N_1], \mathbb{H}_\frak{g}[N_2]\biggr\} + \biggl\{\mathbb{H}_\frak{m}[N_1], \mathbb{H}_\frak{m}[N_2]\biggr\} \notag \\
    &+\biggl[\biggl\{\mathbb{H}_\frak{g}[N_1], \mathbb{H}_\frak{m}[N_2]\biggr\} - \bigl( N_1 \leftrightarrow N_2 \bigr)\biggr].
\end{align}
For readability purposes, we firstly define:
\begin{equation}
    \Delta \biggl\{\mathbb{H}_\frak{g}, \mathbb{H}_\frak{m}\biggr\} \defeq \biggl[\biggl\{\mathbb{H}_\frak{g}[N_1], \mathbb{H}_\frak{m}[N_2]\biggr\} - \bigl( N_1 \leftrightarrow N_2 \bigr)\biggr],
\end{equation}
and
\begin{equation}
    \Delta(\delta N) \defeq \delta N_2 - \delta N_1.
\end{equation}
As before, the fact that $\mathbb{H}_\frak{g}$ (Eq. \ref{eq:HGeo}) does not depend upon the matter sector leads to:
\begin{equation}
    \biggl\{\mathbb{H}_\frak{g}, \mathbb{H}\biggr\}_{\bm{\phi}, \bm{\pi}}=0 \quad \text{ and } \quad \biggl\{\mathbb{H}_\frak{g}, \mathbb{H}\biggr\}_{\delta \phi, \delta \pi}=0.
\end{equation}

Considering the expression of $\mathbb{H}_\frak{m}$ (Eq. \ref{eq:HmDef}), it is clear that the matter sector is basically not depending on the curvature. However, due to the introduction of counter-terms, $\mathbb{H}_\frak{m}$ becomes implicitly a function of $\frak{c}$. On the other hand, $\mathbb{H}_\frak{m}$ is not related to the perturbative expansion of the curvature $\delta K$. It can therefore be concluded that:
\begin{equation}
    \biggl\{\mathbb{H}_\frak{m}, \mathbb{H}_\frak{m}\biggr\}_{\delta E, \delta K}=0.
\end{equation}

These are the main simplifications that can be made {\it a priori} for $\left\{ \mathbb{H}, \mathbb{H} \right\}$. All the other sub-brackets have to be explicitly calculated.
\subsubsection{Bracket $\left\{ \mathbb{H}_\frak{g}, \mathbb{H}_\frak{g} \right\}$}
Let us start with the gravitational part of the Hamiltonian (scalar) constraint $\mathbb{H}_\frak{g}$. For now, we shall focus on the (geometrical) background part of the Poisson bracket:
\begin{subequations}
\begin{align}
\biggl\{\mathbb{H}_\frak{g}[N_1], \mathbb{H}_\frak{g}[N_2]\biggr\}_{\frak{c}, \frak{p}}&=\intwK \mathbf{N} \, \ddN \delta K^b_b \biggl( -g^2 \pc{\tilde{g}+\alpha_1} + 2 g \biggl[ \left[\tilde{g}+\alpha_1\right] \ppp{g} \notag  \\ 
&\qquad+ 2 \frak{p} \biggl( \pp{\tilde{g}+\alpha_1}\, \ppc{g}-\ppp{g}\,\pc{\tilde{g} + \alpha_1}\biggr) \biggr] \biggr)\\
&+ \intwK \frac{\mathbf{N}}{2 \frak{p}} \ddN \delta E^b_b \biggl( - g \left[\tilde{g}^2 + \alpha_2 \right] \ppc{g} - g^2 \tilde{g} \pc{\tilde{g}+\alpha_2} \notag\\ 
&\qquad - 2 g \frak{p} \biggl[ \ppp{\alpha_2} \, \ppc{g} + \ppp{g} \,\ppc{\alpha_2} + 2\tilde{g} [\ppp{g} \ppc{\tilde{g}}-\ppp{\tilde{g}}\ppc{g}] \biggr]\biggr)\\
&+\intwK \frac{\mathbf{N}}{2 \frak{p}} \, \partial^a\bigl(\Delta ( \delta N ) \bigr) \partial_b \bigl( \delta^i_a \delta E^b_i \bigr) \biggl(- g^2 \ppc{\alpha_3}  \notag \\ 
&\qquad - 2 g \biggl[ \left( 1 + \alpha_3 - 2 \frak{p} \ppp{\alpha_3}\right) \ppc{g} + 2 \frak{p} \, \ppp{g} \, \ppc{\alpha_3} \biggr]\biggr).
\end{align}
\end{subequations}
Due to the application of the holonomy correction to the perturbative expansion of $\mathbb{H}_\frak{g}$, the resulting expressions are more complicated than those of \cite{Han:2017wmt}, mostly because of the non-vanishing derivative of $\tilde{g}$. \\

The (geometrical) perturbative phase space bracket reads:
\begin{subequations}
\begin{align}
    \biggl\{\mathbb{H}_\frak{g}[N_1], \mathbb{H}_\frak{g}[N_2]\biggr\}_{\delta E, \delta K}&=\intwK \frac{\mathbf{N}}{2} \ddN \delta K^b_b \biggl( 2 \alpha_2 - \alpha_2 \alpha_4 + 3 \alpha_2 \alpha_5  \notag \\& \qquad - \tilde{g}^2 \bigl[2 + \alpha_4 - 3 \alpha_5 \bigr]  - 4 \alpha_1 \alpha_6 - 4 \tilde{g} \bigl[\alpha_1 + \alpha_6 \bigr] \biggr) \\
    &+ \intwK \frac{\mathbf{N}}{2 \frak{p}} \ddN \delta E^b_b \biggl( 2 \tilde{g}^3 + \alpha_2 \alpha_6 + \tilde{g}^2 \bigl[ \alpha_1 + \alpha_6 \bigr]  \notag \\ 
    &\qquad - 2 \alpha_1 \alpha_7 + 3 \alpha_1 \alpha_8 + \tilde{g} \bigl[ \alpha_2 - 2 \alpha_7 + 3 \alpha_8 \bigr] \biggr) \\
    &+ \intwK \mathbf{N} \, \partial^a \bigl( \ddN \bigr) \partial_a \bigl( \delta K^b_b \bigr) \biggl( \bigl[ 1 + \alpha_3 \bigr] \bigl[ 1 + \alpha_5 \bigr] \biggr) \label{eq:DiffeoDerivTrace} \\
    &+ \intwK \mathbf{N} \, \partial^b \bigl( \ddN \bigr) \partial_a \bigl( \delta^a_i \delta K^i_b \bigr) \biggl( - \bigl[ 1 + \alpha_3 \bigr] \bigl[ 1 + \alpha_4 \bigr] \biggr) \label{eq:DiffeoDerivKDelta} \\
    &+ \intwK \frac{\mathbf{N}}{\frak{p}} \,  \partial^b \bigl( \ddN \bigr) \partial_a \bigl( \delta^i_b \delta E^a_i \bigr) \biggl( \bigl[ 1 + \alpha_3 \bigr] \bigl[ \tilde{g} + \alpha_6 \bigr] \biggr). \label{eq:DiffeoDerivEDelta}
\end{align}
\end{subequations}
In the terms (\ref{eq:DiffeoDerivTrace}), (\ref{eq:DiffeoDerivKDelta}), and (\ref{eq:DiffeoDerivEDelta}), one recognizes the diffeomorphism constraint, Eq.~(\ref{eq:GeoDiffeoExpansionDef}), corrected by the GHC $\tilde{g}$, modulo an additional deformation due to the counter-terms. This comes from the correction applied to the perturbative expansion of the (geometrical) scalar constraint $\mathbb{H}_\frak{g}$.
As for $\left\{ \mathbb{H}, \mathbb{D} \right\}$, this must be accounted for by an anomalous term. Note that, however, such manipulation would not be needed if the (geometrical) diffeomorphism constraint was holonomy-corrected. Thus, we write:
\begin{subequations}
\begin{align}
     \biggl\{\mathbb{H}_\frak{g}[N_1], \mathbb{H}_\frak{g}[N_2]\biggr\}&=\bigl[ 1 + \alpha_3 \bigr] \bigl[ 1 + \alpha_5 \bigr] \, \mathbb{D}_\frak{g} \left[ \frac{\mathbf{N}}{\frak{p}}\partial^a \bigl(\ddN\bigr) \right] \\
     &+\intwK \mathbf{N} \, \ddN \delta K^b_b \, \mathcal{A}_1^{{\left\{ \mathbb{H}, \mathbb{H} \right\}}} \\
     &+ \intwK  \frac{\mathbf{N}}{2 \frak{p}} \, \ddN \delta E^b_b \, \mathcal{A}_2^{{\left\{ \mathbb{H}, \mathbb{H} \right\}}} \\
     &+ \intwK \mathbf{N} \, \partial^b \bigl( \ddN \bigr) \partial_a \bigl( \delta^a_i \delta K^i_b \bigr)  \, \mathcal{A}_3^{{\left\{ \mathbb{H}, \mathbb{H} \right\}}} \\
     &+ \intwK \frac{\mathbf{N}}{\frak{p}} \,  \partial^b \bigl( \ddN \bigr) \partial_a \bigl( \delta^i_b \delta E^a_i \bigr)  \, \mathcal{A}_4^{{\left\{ \mathbb{H}, \mathbb{H} \right\}}},
\end{align}
\end{subequations}
where the first anomalies from the $\left\{ \mathbb{H}, \mathbb{H} \right\}$ bracket are defined as
\begin{align}
    \mathcal{A}_1^{{\left\{ \mathbb{H}, \mathbb{H} \right\}}} &= 2\bigl[ \tilde{g} + \alpha_1 \bigr] \biggl( g \ppc{g} - \tilde{g} - \alpha_6 \biggr)  - \biggl( g^2 + 4 \frak{p} g \ppp{g}\biggr) \pc{\tilde{g}+\alpha_1}  \notag \\
    &\qquad+ 4 \frak{p} g \ppc{g} \pp{\tilde{g}+\alpha_1} + \frac12 \bigl[ \tilde{g}^2 + \alpha_2 \bigr] \biggl( 2 + 3\alpha_5 - \alpha_4 \biggr), \label{eq:AHH1}\\
     \mathcal{A}_2^{{\left\{ \mathbb{H}, \mathbb{H} \right\}}} &= \bigl[ \tilde{g}^2 + \alpha_2 \bigr] \biggl( \tilde{g} + \alpha_6 - g \ppc{g} \biggr) - \frac12 \biggl( g^2 + 4 \frak{p} g \ppp{g} \biggr) \pc{\tilde{g}^2+\alpha_2} \notag\\
      &\qquad+ 2 \frak{p} g \ppc{g} \pp{\tilde{g}^2 + \alpha_2} + \bigl[ \tilde{g}+\alpha_1\bigr]\biggl( \tilde{g}^2 + 3 \alpha_8 - 2 \alpha_7\biggr), \label{eq:AHH2} \\
      \mathcal{A}_3^{{\left\{ \mathbb{H}, \mathbb{H} \right\}}} &= \bigl[ 1 + \alpha_3 \bigr] \bigl[ \alpha_5 - \alpha_4 \bigr], \label{eq:AHH3}\\
       \mathcal{A}_4^{{\left\{ \mathbb{H}, \mathbb{H} \right\}}} &= \bigl[ 1 + \alpha_3 \bigr] \biggl(g\ppc{g} + \bigl[ \tilde{g} + \alpha_6 \bigr] + k \bigl[ 1 + \alpha_5 \bigr] \biggr) - \bigl[ \tilde{g} + \alpha_1 \bigr] \bigl[ 1 + \alpha_9 \bigr] + \notag \\
       &\qquad+ 2 \frak{p} g \ppc{g} \ppp{\alpha_3} + \biggl( \frac12 g^2 + 2 \frak{p} \ppp{g} \biggr) \ppc{\alpha_3}. \label{eq:AHH4}
\end{align}
Those four new terms, $\mathcal{A}_1^{{\left\{ \mathbb{H}, \mathbb{H} \right\}}}, ~ ... ~,\mathcal{A}_4^{{\left\{ \mathbb{H}, \mathbb{H} \right\}}}$ bring the total number of anomalies to 13 at this stage of the derivation. However, some brackets still need to be computed.

\subsubsection{Bracket $\left\{ \mathbb{H}_\frak{m}, \mathbb{H}_\frak{m} \right\}$} 
As discussed in the preceding sections, the (matter) scalar constraint $\mathbb{H}_\frak{m}$ classically does not depend on the extrinsic curvature $K$ and all the Poisson brackets related to the geometrical phase space variables are therefore expected to vanish. However, the counter-terms are functions of the reduced curvature $\frak{c}$. Consequently, $\mathbb{H}_\frak{m}$ is implicitly dependent on $\frak{c}$ as well. The Poisson bracket between matter scalar constraints reads:
\begin{subequations}
\begin{align}
    \biggl\{\mathbb{H}_\frak{m}[N_1], \mathbb{H}_\frak{m}[N_2]\biggr\}_{\frak{c}, \frak{p}}&= \kappa \! \intnK \frac{\mathbf{N} \bm{\pi}}{2\frak{p}} \left(\frac{\bm{\pi}^2}{2 \frak{p}^3} - V[\bm{\phi}]\right) \ddN  \delta \pi \mathcal{A}_5^{{\left\{ \mathbb{H}, \mathbb{H} \right\}}} \\
    &+ \kappa \! \intnK \frac{\mathbf{N}}{2}  \, \partial_\phi \bigl( V[\bm{\phi}] \bigr) \left( \frac{\bm{\pi}^2}{2 \frak{p}} - \frak{p}^2 \, V[\bm{\phi}] \right) \ddN \delta \phi \mathcal{A}_6^{{\left\{ \mathbb{H}, \mathbb{H} \right\}}} \\
    &+ \kappa \! \intnK \frac{\mathbf{N} \bm{\pi}^4}{16 \frak{p}^5} \, \ddN \delta E^b_b \mathcal{A}_7^{{\left\{ \mathbb{H}, \mathbb{H} \right\}}} \\
    &+ \kappa \! \intnK \frac{\mathbf{N} \frak{p}}{4} V[\bm{\phi}]^2 \, \ddN \delta E^b_b \mathcal{A}_8^{{\left\{ \mathbb{H}, \mathbb{H} \right\}}} \\
    &+ \kappa \! \intnK \frac{\mathbf{N} \bm{\pi}^2}{8 \frak{p}^2}  V[\bm{\phi}] \, \ddN \delta E^b_b \mathcal{A}_9^{{\left\{ \mathbb{H}, \mathbb{H} \right\}}},
\end{align}
\end{subequations}
where the five anomalies are given by:
\begin{align}
    \mathcal{A}_5^{{\left\{ \mathbb{H}, \mathbb{H} \right\}}} &= \partial_\frak{c}\beta_1, \label{eq:AHH5}\\
    \mathcal{A}_6^{{\left\{ \mathbb{H}, \mathbb{H} \right\}}} &= \partial_\frak{c}\beta_3, \label{eq:AHH6}\\
    \mathcal{A}_7^{{\left\{ \mathbb{H}, \mathbb{H} \right\}}} &= -\partial_\frak{c}\beta_2, \label{eq:AHH7}\\
    \mathcal{A}_8^{{\left\{ \mathbb{H}, \mathbb{H} \right\}}} &= -\partial_\frak{c}\beta_4, \label{eq:AHH8}\\
    \mathcal{A}_9^{{\left\{ \mathbb{H}, \mathbb{H} \right\}}} &= \pc{\beta_2 + \beta_4}. \label{eq:AHH9}
\end{align}

At this stage, there are 18 anomalies in the list. It can be noticed that, at the background level, only counter-terms related to $\mathbb{H}^{(1)}$ do appear in the anomalies. This is due to the background lapse for $N_1$ and $N_2$ being equal, implying in turn that,
\begin{equation}
    \left\{ \mathbb{H}_\frak{m}^{(2)}[N_1], \mathbb{H}_\frak{m}^{(0)}[N_2] \right\}_{x, p}=\left\{ \mathbb{H}_\frak{m}^{(0)}[N_1], \mathbb{H}_\frak{m}^{(2)}[N_2] \right\}_{x, p},
\end{equation}
where $x$ and $p$ are respectively the canonically conjugate background variables of either the geometrical or matter sectors. This obviously remains true for the background matter sector, whose bracket is given by:
\begin{subequations}
\begin{align}
     \biggl\{\mathbb{H}_\frak{m}[N_1], \mathbb{H}_\frak{m}[N_2]\biggr\}_{\bm{\phi}, \bm{\pi}}&=\intnK \mathbf{N} \, \partial_\phi \bigl( V[\bm{\phi}] \bigr) \, \ddN \delta \pi \biggl( 1 +\beta_1 \biggr) \\
     &+ \intnK \frac{\mathbf{N} \bm{\pi}}{2 \frak{p}} \, \partial_\phi \bigl( V[\bm{\phi}] \bigr) \, \ddN \delta E^b_b \biggl( -2 - \beta_2 - \beta_4\biggr) \\
     &+ \intnK \mathbf{N} \bm{\pi} \,  \partial^2_\phi \bigl( V[\bm{\phi}] \bigr) \, \ddN \delta \phi \biggl( -1 - \beta_3 \biggr).
\end{align}
\end{subequations}
At the level of perturbations, the Poisson bracket is:
\begin{subequations}
\begin{align}
    \biggl\{\mathbb{H}_\frak{m}[N_1], \mathbb{H}_\frak{m}[N_2]\biggr\}_{\delta \phi, \delta \pi}&= \intnK \frac{\mathbf{N} \bm{\pi}}{\frak{p}} \ddN\, \partial_a \partial^a \bigl( \delta \phi \bigr) \biggl( - \bigl[ 1 + \beta_1 \bigr]\bigl[ 1 + \beta_{13} \bigr] \biggr) \label{eq:DiffeoMatterHH} \\
    &+ \intnK \mathbf{N} \, \partial_\phi \bigl( V[\bm{\phi}] \bigr) \, \ddN \delta \pi \biggl( - \bigl[ 1 + \beta_3 \bigr]\bigl[ 1 + \beta_{5} \bigr]  \biggr) \\
    &+ \intnK \mathbf{N} \bm{\pi} \,  \partial^2_\phi \bigl( V[\bm{\phi}] \bigr) \, \ddN \delta \phi \biggl( \bigl[ 1 + \beta_1 \bigr]\bigl[ 1 + \beta_{7} \bigr]  \biggr) \\
    &+ \intnK \frac{\mathbf{N} \bm{\pi}}{2 \frak{p}} \, \partial_\phi \bigl( V[\bm{\phi}] \bigr) \, \ddN \delta E^b_b \biggl( 2 + \beta_1 + \beta_3 + \beta_6 \\
    & \qquad + \beta_3 \beta_6 + \beta_ 8 + \beta_1 \beta_8 \biggr).
\end{align}
\end{subequations}
After integrating by parts, Eq. (\ref{eq:DiffeoMatterHH}) can be shown to be the diffeomorphism constraint associated with the matter sector appearing in Eq. (\ref{eq:DiffeoDensityMatterDef}). The $\left\{ \mathbb{H}_\frak{m}, \mathbb{H}_\frak{m} \right\}$ bracket therefore reads:
\begin{subequations}
\begin{align}
    \biggl\{\mathbb{H}_\frak{m}[N_1], \mathbb{H}_\frak{m}[N_2]\biggr\} &= \bigl[ 1+\beta_1 \bigr]\bigl[ 1 + \beta_{13} \bigr] \, \mathbb{D}_\frak{m}\biggl[\frac{\mathbf{N}}{\frak{p}}\partial^a \bigl(\ddN\bigr) \biggr] \\
    &+ \intnK \mathbf{N} \, \partial_\phi \bigl( V[\bm{\phi}] \bigr) \, \ddN \delta \pi \mathcal{A}_{10}^{{\left\{ \mathbb{H}, \mathbb{H} \right\}}} \\
    &+ \intnK \mathbf{N} \bm{\pi} \,  \partial^2_\phi \bigl( V[\bm{\phi}] \bigr) \, \ddN \delta \phi \mathcal{A}_{11}^{{\left\{ \mathbb{H}, \mathbb{H} \right\}}} \\
    &+ \intnK \frac{\mathbf{N} \bm{\pi}}{2 \frak{p}} \, \partial_\phi \bigl( V[\bm{\phi}] \bigr) \, \ddN \delta E^b_b \mathcal{A}_{12}^{{\left\{ \mathbb{H}, \mathbb{H} \right\}}} \\
     &+ \biggl\{\mathbb{H}_\frak{m}[N_1], \mathbb{H}_\frak{m}[N_2]\biggr\}_{\frak{c}, \frak{p}},
\end{align}
\end{subequations}
giving rise to 3 new anomalies, 
\begin{align}
    \mathcal{A}_{10}^{{\left\{ \mathbb{H}, \mathbb{H} \right\}}} &= \beta_1 - \beta_3 - \beta_5 - \beta_3 \beta_5, \label{eq:AHH10}\\
    \mathcal{A}_{11}^{{\left\{ \mathbb{H}, \mathbb{H} \right\}}} &= \beta_1 + \beta_7 + \beta_1 \beta_7 - \beta_3, \label{eq:AHH11}\\
    \mathcal{A}_{12}^{{\left\{ \mathbb{H}, \mathbb{H} \right\}}} &= \beta_1 + \beta_3 + \beta_6 + \beta_3 \beta_6 + \beta_8 + \beta_1 \beta_8 - \beta_2 - \beta_4, \label{eq:AHH12}
\end{align}
bringing the total amount to 21 terms.
\subsubsection{Bracket $\left\{ \mathbb{H}_\frak{g}, \mathbb{H}_\frak{m} \right\}$}
A single bracket remains to be calculated.
Given the above computations of $\left\{ \mathbb{H}_\frak{g}, \mathbb{H}_\frak{g} \right\}$ and $\left\{ \mathbb{H}_\frak{m}, \mathbb{H}_\frak{m} \right\}$, where the expected result is recovered -- although deformed by a combination of counter-terms -- we expect $\left\{ \mathbb{H}_\frak{g}, \mathbb{H}_\frak{m} \right\}$ to be purely anomalous.\\

The background part of the bracket is
\begin{subequations}
\begin{align}
     \Delta \biggl\{\mathbb{H}_\frak{g}, \mathbb{H}_\frak{m}\biggr\}_{\frak{c}, \frak{p}} &= \intnK \frac{\mathbf{N} \bm{\pi}^2}{ 2 \frak{p}^2} \, \ddN \delta K^b_b \pc{\tilde{g}+ \alpha_1} \\
    &+ \intnK \frac{\mathbf{N} \frak{p}}{2} V[\bm{\phi}] \, \ddN \delta K^b_b \pc{-\tilde{g} - \alpha_1} \\
    &+ \intnK \frac{\mathbf{N}}{2} \biggl( \frac{\bm{\pi}^2}{2 \frak{p}^3} - V[\bm{\phi}] \biggr) \partial^a \bigl( \ddN \bigr) \partial_b \bigl( \delta^i_a \delta E^b_i \bigr) \ppc{\alpha_3} \\
    &+ \intnK \frac{\mathbf{N} \bm{\pi}}{2 \frak{p}^2} \, \ddN \delta \pi \biggl( 2g \ppc{g} \bigl( 3 + 3 \beta_1 - 2 \frak{p} \ppp{\beta_1} \bigr) + g^2 \ppc{\beta_1} \notag \\
    &\qquad + 4 g \frak{p} \ppp{g} \ppc{\beta_1} \biggr) \\
    &+ \intnK \frac{\mathbf{N} \bm{\pi}^2}{8 \frak{p}^3} \ddN \delta E^b_b \biggl( 2 \tilde{g} \ppc{\tilde{g}}  - \ppc{\alpha_2} + g^2 \ppc{\beta_2} + 4 g \frak{p} \ppp{g} \ppc{\beta_2} \biggr)  \notag\\
    &\qquad + 2 g \ppc{g} \bigl( 5 + 5 \beta_2 - 2 \frak{p} \ppp{\beta_2} \bigr) \\
    &+ \intnK \frac{\mathbf{N}}{4} V[\bm{\phi}] \, \ddN \delta E^b_b \biggl( - 2 g \ppc{g} \bigl( 1 + \beta_4 + 2 \frak{p} \ppp{\beta_4} \bigr) + g^2 \ppc{\beta_4} \notag \\
    &\qquad+ \ppc{\alpha_2} + 4 g \frak{p} \ppp{g} \ppc{\beta_4} \biggr) \\
    &+ \intnK \mathbf{N} \, \partial_\phi \bigl( V[\bm{\phi}] \bigr) \ddN \delta \phi \biggl( - g \ppc{g} \bigl( 3 \frak{p} \bigl[ 1 + \beta_3 \bigr]+ 2 \frak{p}^2 \ppp{\beta_3} \bigr) \notag\\
    &\qquad + \frac{\frak{p}}{2} g^2 \ppc{\beta_3}  + 2 \frak{p}^2 g \ppp{g} \ppc{\beta_3} \biggr),
\end{align}
\end{subequations}
while it becomes for perturbations:
\begin{subequations}
\begin{align}
      \Delta \biggl\{\mathbb{H}_\frak{g}, \mathbb{H}_\frak{m}\biggr\}_{\delta E, \delta K} &= \intnK \frac{\mathbf{N} \bm{\pi}^2}{ 2 \frak{p}^2} \, \ddN \delta K^b_b \biggl( \bigl[ 1 + \beta_2 \bigr]\bigl( 2 - \alpha_4 + 3 \alpha_5 \bigr) \biggr) \\
     &+  \intnK \frac{\mathbf{N} \frak{p}}{2} V[\bm{\phi}] \, \ddN \delta K^b_b \biggl(  \bigl[ 1 + \beta_4 \bigr]\bigl( -2 + \alpha_4 - 3 \alpha_5 \bigr) \biggr) \\
     &+  \intnK  \frac{\mathbf{N} \bm{\pi}}{2 \frak{p}^2} \, \ddN \delta \pi \biggl( - 6\bigl[ \tilde{g}+ \alpha_1 \bigr] \bigl[ 1 + \beta_6 \bigr] \biggr) \\
     &+  \intnK \frac{\mathbf{N} \bm{\pi}^2}{4 \frak{p}^3} \ddN \delta E^b_b \biggl( \alpha_6 \bigl[  1 + \beta_2 \bigr] + \alpha_1 \bigr( 5 + 2 \beta_{10} + 3 \beta_9\bigr) \notag \\
     &\qquad+ \tilde{g} \bigl( 6 + 2 \beta_{10} + \beta_2 + 3 \beta_9 \bigr) \biggr) \\
     &+ \intnK \frac{\mathbf{N}}{4} V[\bm{\phi}] \, \ddN \delta E^b_b \biggl( \alpha_1 \bigl(-3 \beta_{11}+2 \beta_{12}-1\bigr)+\alpha_6 \bigl[1+\beta_4\bigr]\notag \\
     &\qquad+\tilde{g} (-3\beta_{11}+2 \beta_{12}+\beta_4) \biggr)  \\
     &+ \intnK \mathbf{N} \, \partial_\phi \bigl( V[\bm{\phi}] \bigr) \ddN \delta \phi \biggl( 3 \frak{p} \bigl[ \tilde{g} + \alpha_1 \bigr] + 3 \frak{p} \bigl[\tilde{g}+\alpha_1\bigr] \beta_8 \biggr).
\end{align} 
\end{subequations}
This leads to:
\begin{subequations}
\begin{align}
      \Delta \biggl\{\mathbb{H}_\frak{g}, \mathbb{H}_\frak{m}\biggr\} &= \intnK \frac{\mathbf{N} \bm{\pi}^2}{ 4 \frak{p}^2} \, \ddN \delta K^b_b \mathcal{A}_{13}^{{\left\{ \mathbb{H}, \mathbb{H} \right\}}} \\
     &+ \intnK \frac{\mathbf{N} \frak{p}}{2} V[\bm{\phi}] \, \ddN \delta K^b_b \mathcal{A}_{14}^{{\left\{ \mathbb{H}, \mathbb{H} \right\}}} \\
     &+ \intnK \frac{\mathbf{N}}{2} \biggl( \frac{\bm{\pi}^2}{2 \frak{p}^3} - V[\bm{\phi}] \biggr) \partial^a \bigl( \ddN \bigr) \partial_b \bigl( \delta^i_a \delta E^b_i \bigr) \mathcal{A}_{15}^{{\left\{ \mathbb{H}, \mathbb{H} \right\}}} \\
     &+ \intnK \frac{\mathbf{N} \bm{\pi}}{2 \frak{p}^2} \, \ddN \delta \pi \mathcal{A}_{16}^{{\left\{ \mathbb{H}, \mathbb{H} \right\}}} \\
     &+ \intnK \frac{\mathbf{N} \bm{\pi}^2}{8 \frak{p}^3} \ddN \delta E^b_b \mathcal{A}_{17}^{{\left\{ \mathbb{H}, \mathbb{H} \right\}}} \\
     &+ \intnK \frac{\mathbf{N}}{4} V[\bm{\phi}] \, \ddN \delta E^b_b \mathcal{A}_{18}^{{\left\{ \mathbb{H}, \mathbb{H} \right\}}} \\
     &+ \intnK \mathbf{N} \, \partial_\phi \bigl( V[\bm{\phi}] \bigr) \ddN \delta \phi \mathcal{A}_{19}^{{\left\{ \mathbb{H}, \mathbb{H} \right\}}},
\end{align}
\end{subequations}
where seven new anomalies appear:
\begin{align}
    \mathcal{A}_{13}^{{\left\{ \mathbb{H}, \mathbb{H} \right\}}} &=  \bigl[ 1 + \beta_2 \bigr]\bigl( 2 - \alpha_4 + 3 \alpha_5 \bigr) - 2 \pc{\tilde{g}+ \alpha_1}, \label{eq:AHH13}\\
    \mathcal{A}_{14}^{{\left\{ \mathbb{H}, \mathbb{H} \right\}}} &= \bigl[ 1 + \beta_4 \bigr]\bigl( -2 + \alpha_4 - 3 \alpha_5 \bigr) + 2 \pc{\tilde{g}+ \alpha_1} \label{eq:AHH14}\\
    \mathcal{A}_{15}^{{\left\{ \mathbb{H}, \mathbb{H} \right\}}} &= \ppc{\alpha_3} \label{eq:AHH15}\\
    \mathcal{A}_{16}^{{\left\{ \mathbb{H}, \mathbb{H} \right\}}} &= - 6\bigl[ \tilde{g}+ \alpha_1 \bigr] \bigl[ 1 + \beta_6 \bigr] + 2g \ppc{g} \bigl( 3 + 3 \beta_1 - 2 \frak{p} \ppp{\beta_1} \bigr) + g^2 \ppc{\beta_1} + 4 g \frak{p} \ppp{g} \ppc{\beta_1}, \label{eq:AHH16} \\
    \mathcal{A}_{17}^{{\left\{ \mathbb{H}, \mathbb{H} \right\}}} &= 12 \tilde{g} + 10 \alpha_1 + 2 \alpha_6 + 4 \bigl[\tilde{g}+\alpha_1\bigr] \beta_{10} + 2 \bigl[\tilde{g}+\alpha_6\bigr] \beta_2  + 6 \tilde{g} \beta_9 - 2 \tilde{g} \ppc{\tilde{g}} \notag\\ 
    &\qquad - 2 g \ppc{g} \bigl( 5 + 5 \beta_2 - 2 \frak{p} \ppp{\beta_2} \bigr)  - \ppc{\alpha_2} - g^2 \ppc{\beta_2} - 4 g \frak{p} \ppp{g} \ppc{\beta_2},\label{eq:AHH17}\\
    \mathcal{A}_{18}^{{\left\{ \mathbb{H}, \mathbb{H} \right\}}} &= 2 \alpha_1 - 2 \alpha_6 + 6 \bigl[\tilde{g} + \alpha_1 \bigr] \beta_{11} - 4 \bigl[\tilde{g}+\alpha_1\bigr] \beta_{12} - 2 \bigl[\tilde{g} + \alpha_6 \bigr]\beta_4 + 2 \tilde{g} \ppc{\tilde{g}} \notag \\
    &\qquad - 2 g \ppc{g} \bigl( 1 + \beta_4 + 2 \frak{p} \ppp{\beta_4} \bigr) + \ppc{\alpha_2} + g^2 \ppc{\beta_4} + 4 g \frak{p} \ppp{g} \ppc{\beta_4}, \label{eq:AHH18}\\
    \mathcal{A}_{19}^{{\left\{ \mathbb{H}, \mathbb{H} \right\}}} &= 3 \frak{p} \bigl[ \tilde{g} + \alpha_1 \bigr] + 3 \frak{p} \bigl[\tilde{g}+\alpha_1\bigr] \beta_8- g \ppc{g} \bigl( 3 \frak{p} \bigl[ 1 + \beta_3 \bigr] + 2 \frak{p}^2 \ppp{\beta_3} \bigr) + \frac{\frak{p}}{2} g^2 \ppc{\beta_3}\notag\\
    &\qquad  + 2 \frak{p}^2 g \ppp{g} \ppc{\beta_3}. \label{eq:AHH19}
\end{align}

The previous Poisson bracket and associated anomalies complete the calculation of the full algebra in the presence of generalized holonomy corrections, applied either to the background or to the perturbative expansion of the scalar constraint. The next step consists in cancelling those anomalies.

%% file: texts/solution_algebra/solution_ct.tex
The counter-terms must now be determined so that the anomalies do vanish. The procedure we follow, initiated in particular in \cite{Bojowald:2008gz,Mielczarek:2011ph}, is close to the one described recently in \cite{Han:2017wmt}. However, as the holonomy corrections are here applied not only on the background part of the scalar constraint, but also on the perturbative expansion, the calculation needs to be explicitely performed to investigate possible new closure conditions.\\

Some anomalies are related to a single counter-term and can therefore be trivially cancelled with
\begin{equation}
    \beta_4 = 0,\quad \beta_8 = 0,\quad \beta_{10} = 0, \quad \text{ and } \quad \beta_{12}=0,
\end{equation}
from, respectively, $\mathcal{A}_8^{{\left\{ \mathbb{H}, \mathbb{D} \right\}}}$ (\ref{eq:AHD8}), $\mathcal{A}_{9}^{{\left\{ \mathbb{H}, \mathbb{D} \right\}}}$ (\ref{eq:AHD9}), $\mathcal{A}_{10}^{{\left\{ \mathbb{H}, \mathbb{D} \right\}}}$ (\ref{eq:AHD10}) and $\mathcal{A}_{12}^{{\left\{ \mathbb{H}, \mathbb{D} \right\}}}$ (\ref{eq:AHD12}). In a way, this just means that the associated counter-terms were not necessary from the beginning. Moreover, combining $\beta_{12}=0$ and  $\mathcal{A}_{13}^{{\left\{ \mathbb{H}, \mathbb{D} \right\}}}$ (\ref{eq:AHD13}), one obtains:
\begin{equation}
    \beta_{11}=0.
\end{equation}
Since $\beta_4=0$, it can be concluded from $\mathcal{A}_{14}^{{\left\{ \mathbb{H}, \mathbb{H} \right\}}} = 0$ (\ref{eq:AHH14}) that,
\begin{equation}
     \pc{\tilde{g}+ \alpha_1} = \frac12 \bigl( 2 - \alpha_4 + 3 \alpha_5 \bigr), \label{eq:DerivA45}
\end{equation}
which implies, due to $\mathcal{A}_{13}^{{\left\{ \mathbb{H}, \mathbb{H} \right\}}} = 0$ (\ref{eq:AHH13}),
\begin{equation}
   \beta_2 \bigl( 2 - \alpha_4 + 3 \alpha_5 \bigr)=0.
\label{beta2 alpha4 alpha5 equation}
\end{equation}
As we shall see later in this section, the cancellation of $\mathcal{A}_{3}^{{\left\{ \mathbb{H}, \mathbb{H} \right\}}}$ (\ref{eq:AHH3}) requires $\alpha_4=\alpha_5$. If $\beta_2\neq 0$ is assumed, the previous equality implies $\alpha_4=1$, which spoils the classical limit of the counter-term. As a consequence, Eq. (\ref{beta2 alpha4 alpha5 equation}) implies
\begin{equation}
    \beta_2 = 0.
\end{equation}
It should be noticed that this cancellation does not depend on whether corrections have been implemented on the perturbative expansion of the constraints or not. Considering both $\beta_2=0$ and  $\mathcal{A}_7^{{\left\{ \mathbb{H}, \mathbb{D} \right\}}}=0$ (\ref{eq:AHD7}), one gets:
\begin{equation}
    \beta_1 = 0.
\end{equation}
From $\mathcal{A}_{12}^{{\left\{ \mathbb{H}, \mathbb{H} \right\}}}=0$ (\ref{eq:AHH12}), the following relation can be derived:
\begin{equation}
    \beta_6 = - \frac{\beta_3}{1+\beta_3},
\end{equation}
such that $\mathcal{A}_{16}^{{\left\{ \mathbb{H}, \mathbb{H} \right\}}}=0$ (\ref{eq:AHH16}) and $\beta_1=0$ imply
\begin{equation}
     \beta_3 = \frac{\alpha_1 + \tilde{g}}{g \, \ppc{g}} - 1. \label{eq:Alpha1}
\end{equation}
Given this expression, the cancellation of $\mathcal{A}_{19}^{{\left\{ \mathbb{H}, \mathbb{H} \right\}}}$ (\ref{eq:AHH19}) and $\mathcal{A}_{6}^{{\left\{ \mathbb{H}, \mathbb{H} \right\}}}$ (\ref{eq:AHH6}) implies:
\begin{equation}
    \ppp{\beta_3}=0. \label{eq:IndePB3}
\end{equation}
To ensure the correct classical limit (i.e. $\frak{c}\rightarrow0$), counter-terms cannot be independent of the non-trivial part of the extrinsic curvature $\frak{c}$ -- via $\mathcal{A}_{6}^{{\left\{ \mathbb{H}, \mathbb{H} \right\}}}$ (\ref{eq:AHH6}) -- \textit{and} of the non-trivial part of the densitized triads $\frak{p}$, see Eq. (\ref{eq:IndePB3}). This leads to:
\begin{equation}
    \beta_3 = 0 \quad \text{ and } \quad \beta_6=0.
\end{equation}
Still, Eq. (\ref{eq:Alpha1}) must hold with $\beta_3=0$, which implies:
\begin{equation}
    \alpha_1 = g \ppc{g} - \tilde{g}. \label{eq:Alpha1Def}
\end{equation}
The functional dependence of this particular counter-term, $\alpha_1$, depends on whether the holonomy correction has been applied to the perturbed part of the scalar constraint or not. \\

The cancellation of $\mathcal{A}_{6}^{{\left\{ \mathbb{H}, \mathbb{D} \right\}}}$ (\ref{eq:AHD6}), $\mathcal{A}_{11}^{{\left\{ \mathbb{H}, \mathbb{D} \right\}}}$ (\ref{eq:AHD11}), and $\mathcal{A}_{11}^{{\left\{ \mathbb{H}, \mathbb{H} \right\}}}$ (\ref{eq:AHH11}) implies
\begin{equation}
    \beta_5=0, \quad \beta_9=0, \quad \text{ and } \quad \beta_7=0,
\end{equation}
such that only $\beta_{13}$ remains undetermined for the matter sector. In particular, $\beta_{13}$ does not appear in any anomaly and will be fixed later by requiring to recover the expected, although deformed, structure of the algebra of constraint for the $\left\{ \mathbb{H}, \mathbb{H}\right\}$ bracket, namely $\left\lbrace \mathbb{H}, \mathbb{H} \right\rbrace \propto \mathbb{D}$. \\ 

Considering $\mathcal{A}_{1}^{{\left\{ \mathbb{H}, \mathbb{D} \right\}}}=0$ (\ref{eq:AHD1}) together with Eq. (\ref{eq:Alpha1Def}) leads to
\begin{equation}
    \alpha_2 = 3 g^2 - \tilde{g}^2 - 2 \frak{c} g \ppc{g}. \label{eq:Alpha2Def}
\end{equation}
Again, this expression for $\alpha_2$ does depend on whether the GHC has been applied to the perturbative expansion of the geometrical scalar constraint or not. Since $\alpha_3$ cannot be a constant (so as to have the correct behavior in the classical limit), it can be concluded from $\mathcal{A}_{3}^{{\left\{ \mathbb{H}, \mathbb{H} \right\}}}=0$~(\ref{eq:AHH3}) that:
\begin{equation}
    \alpha_5 = \alpha_4.
\end{equation}
Given Eq. (\ref{eq:DerivA45}), this implies
\begin{equation}
    \alpha_4 = \bigl(\ppc{g}\bigr)^2 + g \partial_\frak{c}^2 g - 1. \label{eq:Alpha4Def}
\end{equation}
Since $\alpha_4$ and $\alpha_5$ are known,  $\mathcal{A}_{3}^{{\left\{ \mathbb{H}, \mathbb{D} \right\}}}$ (\ref{eq:AHD3}) or $\mathcal{A}_{5}^{{\left\{ \mathbb{H}, \mathbb{D} \right\}}}$ (\ref{eq:AHD5}) can be cancelled with:
\begin{equation}
    \alpha_6=2 g \ppc{g} - \tilde{g} - \frak{c} \bigl[ \bigl(\ppc{g}\bigr)^2 + g \partial_\frak{c}^2 g \bigr]. \label{eq:Alpha6Def}
\end{equation}
Solving $\mathcal{A}_{2}^{{\left\{ \mathbb{H}, \mathbb{D} \right\}}}=0$ (\ref{eq:AHD2}), this means that
\begin{equation}
    \alpha_7 = \alpha_8.
\end{equation}
Considering $\mathcal{A}_{4}^{{\left\{ \mathbb{H}, \mathbb{D} \right\}}}=0$ (\ref{eq:AHD4}), one then gets
\begin{equation}
    \alpha_7 = 2 \frak{c} \biggl( 2 g \ppc{g} - \frak{c} \bigl(\ppc{g}\bigr)^2 - \frak{c} g \partial_\frak{c}^2 g \biggr) - 4 \frak{p} g \ppp{g} - g^2 - \tilde{g}^2.
\end{equation}
Substituting Eqs. (\ref{eq:Alpha1Def}), (\ref{eq:Alpha4Def}), and (\ref{eq:Alpha6Def}) into $\mathcal{A}_{4}^{{\left\{ \mathbb{H}, \mathbb{H} \right\}}}=0$ (\ref{eq:AHH4}) and making use of $\mathcal{A}_{14}^{{\left\{ \mathbb{H}, \mathbb{H} \right\}}} =0$ (\ref{eq:AHH14}) leads to a relation between $\alpha_3$ and $\alpha_9$:
\begin{equation}
    g \ppc{g} \biggl[ \alpha_3 + 2 \frak{p} \ppp{\alpha_3} - \alpha_9 \biggr] = 0.
\end{equation}
This exhibits a freedom in the choice of $\alpha_3$ and $\alpha_9$. This degeneracy is often overlooked in the literature. In \cite{Han:2017wmt}, the choice $\alpha_3=0$ and $\alpha_9=0$ was made so as to ensure the compatibility with results obtained in curved space. However at this stage, it seems that they were derived with inverse-volume corrections \textit{only} \cite{Han:2018usc}. It looks like to us that the question therefore remains quite open and, in this work, we choose to remain as generic as possible. \\

Considering $\mathcal{A}_{15}^{{\left\{ \mathbb{H}, \mathbb{H} \right\}}}$ (\ref{eq:AHH15}), one obtains:
\begin{equation}
    \alpha_3 \defeq f(\frak{p}) = \frac{1}{\sqrt{\frak{p}}}\biggl[K + \int_1^{\frak{p}} \! \! d \Theta \frac{\alpha_9(\Theta)}{\sqrt{\Theta}} \biggr], \label{eq:Alpha3Def}
\end{equation}
where $K$ is some constant. \\

The only term which remains to be fixed is now $\beta_{13}$. It can be determined by requiring the $\left\{ \mathbb{H}, \mathbb{H} \right\}$ bracket to be proportional to the \textit{full} diffeomorphism constraint, implying that the structure coefficients in front of $\mathbb{D}_{\frak{g}}$ and $\mathbb{D}_{\frak{m}}$ are the same.
This translates, using Eqs. (\ref{eq:Alpha3Def}) and (\ref{eq:Alpha4Def}), in:
\begin{equation}
    \beta_{13} = \pc{g \ppc{g}} \biggl[ 1 + f(\frak{p}) \biggr] - 1. \label{eq:beta13}
\end{equation}
This concludes the calculation of all the counter-terms. \\

It is however important to check that the remaining anomalies are solved using the counter-terms that have just been derived. First of all, $\mathcal{A}_{1}^{{\left\{ \mathbb{H}, \mathbb{H} \right\}}}$ (\ref{eq:AHH1}) leads to a condition on $g$:
\begin{equation}
    g - 2 \frak{p} \ppp{g} - \frak{c} \ppc{g} = 0.
\end{equation}
Interestingly, this is the same as the one derived in \cite{Han:2017wmt} where holonomy corrections were applied only to the background part of the scalar constraint and not on the perturbed expansion. Although $\mathcal{A}_{1}^{{\left\{ \mathbb{H}, \mathbb{H} \right\}}}$ depends explicitly, and implicitly via the counter-terms, on the GHC $\tilde{g}$ implemented at the perturbative level, the calculations 
show that no additional restriction appears on $\tilde{g}$ when compared to the case where only the background is corrected. \\

In addition, it is easy to check that, given all counter-terms derived from the matter sector\footnote{Each one vanishes except $\beta_{13}$, see Eq. (\ref{eq:beta13}).}, all the remaining anomalies which depend only on the matter sector counter-terms are automatically cancelled. Otherwise stated, $\mathcal{A}_{5}^{{\left\{ \mathbb{H}, \mathbb{H} \right\}}}$ (\ref{eq:AHH5}), $\mathcal{A}_{7}^{{\left\{ \mathbb{H}, \mathbb{H} \right\}}}$ (\ref{eq:AHH7}), $\mathcal{A}_{8}^{{\left\{ \mathbb{H}, \mathbb{H} \right\}}}$ (\ref{eq:AHH8}), $\mathcal{A}_{9}^{{\left\{ \mathbb{H}, \mathbb{H} \right\}}}$ (\ref{eq:AHH9}), and $\mathcal{A}_{10}^{{\left\{ \mathbb{H}, \mathbb{H} \right\}}}$ (\ref{eq:AHH10}) are  trivially set to zero with the provided solution. Finally, the same argument applied to $\mathcal{A}_{17}^{{\left\{ \mathbb{H}, \mathbb{H} \right\}}}$ (\ref{eq:AHH17}) and $\mathcal{A}_{18}^{{\left\{ \mathbb{H}, \mathbb{H} \right\}}}$ (\ref{eq:AHH18}) leads to the consistency conditions
\begin{equation}
    12 \tilde{g} + 10 \alpha_1 + 2 \alpha_6 - 10 g \ppc{g} - 2 \tilde{g} \ppc{\tilde{g}} - \ppc{\alpha_2} = 0,
\end{equation}
and
\begin{equation}
    2 \alpha_1 - 2 \alpha_6 + 2 \tilde{g} \ppc{\tilde{g}} - 2 g \ppc{g} + \ppc{\alpha_2} = 0,
\end{equation}
which were already satisfied given Eqs. (\ref{eq:Alpha1Def}), (\ref{eq:Alpha2Def}), and (\ref{eq:Alpha6Def}). \\

This completes the procedure to ensure the closing of the algebra of constraints. The resulting structure is deformed with respect to GR, as shown by Eq. (\ref{eq:beta13}). 

To summarize, we have:
\begin{align}
    \bigl\{\mathbb{G}, \mathbb{G}\bigr\} &= 0, \\
    \bigl\{\mathbb{D}, \mathbb{D}\bigr\} &= 0, \\
    \bigl\{\mathbb{D}, \mathbb{G}\bigr\} &= 0, \\
    \bigl\{\mathbb{H}, \mathbb{G}\bigr\} &= 0,
\end{align}
\begin{equation}
    \biggl\{\mathbb{H}[N], \mathbb{D}[N^a]\biggr\} = \mathbb{H}[\delta N \partial_a \delta N^a],
\end{equation}  
\begin{equation}
    \biggl\{\mathbb{H}[N_1], \mathbb{H}[N_2]\biggr\} = \mathcal{G}^{(2)} \bigl[ 1 + f(\frak{p}) \bigr] \mathbb{D}\biggl[\frac{\mathbf{N}}{\frak{p}}\partial^a \bigl(\ddN\bigr) \biggr],
    \label{alg}
\end{equation}
where we have used the same notation than in \cite{Han:2017wmt}, i.e.
\begin{equation}
    \mathcal{G}^{(2)} \defeq \mathcal{G}^{(2)}(\frak{c}, \frak{p}) \defeq \frac12 \partial_\frak{c}^2 g^2.
\end{equation}

%% file: texts/solution_algebra/solution_restriction_ghc.tex
As discussed at the beginning of this article, it is interesting to wonder whether the \textit{a priori} freedom that exists in choosing the shape of the GHC can be used to close of the algebra without any counter-term. It has been nicely shown in \cite{Li:2023axl} that, focusing only on the vector modes of the cosmological perturbations \cite{Lifshitz:1945du}, it is indeed possible to form a first-class algebra. We argue in this section that when scalar modes are considered as well, this interesting conclusion unfortunately does not hold anymore. \\

Using the properties of  vector modes \cite{Lifshitz:1945du}, namely, $\delta N = 0$, $\delta E^b_b = 0$, $\delta K^b_b=0$, and $\mathcal{Z}_{ab}^{cidj}=0$, it was shown \cite{Li:2023axl} that\footnote{More precisely, in \cite{Li:2023axl}, the choice was made to apply the GHC at the perturbative level of the Hamiltonian constraint $\mathbb{H}^{|v}_\frak{g}$, that is to say $\Tilde{g}=g$ following the notation of this paper. Although it does not change the conclusion, we prefer to keep this subtlety visible when presenting the results to be fully consistent.}:
\begin{subequations}
\label{eq:DVVec}
\begin{align}
	\biggl\{\mathbb{H}^{|v}_\frak{g}[N], \mathbb{D}^{|v}_\frak{g}[N^a] \biggr\} &= \frak{D} \, \mathbb{D}^{|v}_\frak{g}\biggl[\frac{\mathbf{N}}{\sqrt{\frak{p}}} \delta N^a\biggr] \label{eq:DVVec1}\\
    &+ \intwK \frac{\mathbf{N}}{2\sqrt{\frak{p}}} \,\partial_a \bigl(\delta N^b\bigr) \delta^i_b \delta E^a_i \mathcal{A}_{v}^{{\left\{ \mathbb{H}, \mathbb{D} \right\}}} \label{eq:DVVec2},
\end{align}
\end{subequations}
where $X^{|v}$ means that properties of vector modes have been applied to the quantity $X$, whereas $\frak{D}$ is a deformation coefficient defined as,
\begin{equation}
    \frak{D} \defeq \frak{c} + \tilde{g} - 2 g \ppc{g},
\end{equation} 
and $\mathcal{A}_{v}^{{\left\{ \mathbb{H}, \mathbb{D} \right\}}}$ is given by:
\begin{equation}
    \mathcal{A}_{v}^{{\left\{ \mathbb{H}, \mathbb{D} \right\}}} = \frac{\tilde{g}^2}{2} + \frac{g^2}{2} - \frak{c}^2 - 2\frak{c}\tilde{g} + 2 g \biggl( \frak{p} \ppp{g} + \frak{c} \ppc{g} \biggr).
    \label{eq:Avec}
\end{equation}
Since $\delta N=0$ for vector modes, the Poisson bracket (\ref{eq:DVVec}) is the only relevant one in that case. Thus, the cancellation of the previous anomaly $\mathcal{A}_{v}^{{\left\{ \mathbb{H}, \mathbb{D} \right\}}}$ (\ref{eq:Avec}) needed to ensure that the algebra of constraints remains first-class imposes a specific restriction on the GHC. The paramount importance of those restrictions for phenomenology was already underlined \cite{Renevey:2021tmh, DeSousa:2022rep,Li:2023axl}. Using the detailed calculations presented in the previous sections, and forcing counter-terms to vanish (as the algebra is here closed only by the GHC), it can be seen that Eq. (\ref{eq:DVVec}) and Eq. (\ref{eq:HDFull}) are equivalent once symmetries have been applied:
\begin{subequations}
\begin{align}
    \biggl\{\mathbb{H}^{|v}_\frak{g}[N], \mathbb{D}^{|v}_\frak{g}[N^a] \biggr\} &= \intwK \frac{\mathbf{N}}{2\sqrt{\frak{p}}} \,\partial_a \bigl(\delta N^b\bigr) \delta^i_b \delta E^a_i \mathcal{A}_4^{{\left\{ \mathbb{H}, \mathbb{D} \right\}}} \\
     &+ \intwK \mathbf{N} \sqrt{\frak{p}} \,\partial_a \bigl(\delta N^b\bigr) \delta^a_i \delta K^i_b  \mathcal{A}_5^{{\left\{ \mathbb{H}, \mathbb{D} \right\}}}.
\end{align}
\end{subequations}

Although enforcing the algebra of constraints to be first-class is a strong requirement, ambiguities remain, even in this simple situation. In particular, when focusing on vector modes only, two different scenarios lead to a closed algebra, resulting in different restrictions on the GHC. First, as discussed in \cite{Li:2023axl}, the restriction $\mathcal{A}_{v}^{{\left\{ \mathbb{H}, \mathbb{D} \right\}}} = 0$ (\ref{eq:Avec}) leads to a satisfying solution. However, the restrictions $\mathcal{A}_4^{{\left\{ \mathbb{H}, \mathbb{D} \right\}}} = 0$ (\ref{eq:AHD4}) \textit{and} $\mathcal{A}_5^{{\left\{ \mathbb{H}, \mathbb{D} \right\}}} = 0$ (\ref{eq:AHD5}) also result in a first-class algebra. Note, however, that the solution space for $g(\frak{c}, \frak{p})$ under the condition $\mathcal{A}_{v}^{{\left\{ \mathbb{H}, \mathbb{D} \right\}}} = 0$ (\ref{eq:Avec}) is less restrictive than the one for $\mathcal{A}_4^{{\left\{ \mathbb{H}, \mathbb{D} \right\}}} = 0$ (\ref{eq:AHD4}) and $\mathcal{A}_5^{{\left\{ \mathbb{H}, \mathbb{D} \right\}}} = 0$ (\ref{eq:AHD5}), as the latter condition implies $\mathcal{A}_{v}^{{\left\{ \mathbb{H}, \mathbb{D} \right\}}} = 0$ (\ref{eq:Avec}) by construction.  Otherwise stated, instead of forcing the appearance of the diffeomorphism constraint, one could simply ensure that the bracket vanishes. These ambiguities have to be explicitly addressed to extend the results of \cite{Li:2023axl} to scalar modes. This is the aim of the following of this section. \\

Starting with the $\left\{ \mathbb{H}, \mathbb{D} \right\}$ bracket, the above-mentioned ambiguity is elegantly fixed by the introduction of matter. The latter is  \textit{mandatory} for studying scalar modes. As soon as matter is considered, the diffeomorphism constraint $\mathbb{D}$ is given by $ \mathbb{D}=\mathbb{D}_\frak{g} + \mathbb{D}_\frak{m}$. Thus, to obtain\footnote{We derived in the previous section that, if scalar modes are considered, the $\left\{ \mathbb{H}, \mathbb{D} \right\}$ bracket is at least proportional to the scalar constraint $\mathbb{H}$, as expected from classical considerations.} $\left\{ \mathbb{H}, \mathbb{D} \right\} \supset \mathbb{D}$, one has to investigate the $\left\{ \mathbb{H}_\frak{m}, \mathbb{D} \right\}$ bracket. This was done in the previous section: without counter-terms, Eq. (\ref{eq:HDtot}) implies that $\left\{ \mathbb{H}_\frak{m}, \mathbb{D} \right\}=\mathbb{H}_\frak{m}$, ensuring that no anomaly emerges from this particular sub-bracket. This means that it is not possible to obtain $\left\{ \mathbb{H}, \mathbb{D} \right\} \supset \mathbb{D}$ while remaining consistent. The ambiguity discussed above is therefore fixed by the introduction of matter contributions.\\

As for the $\left\{ \mathbb{H}, \mathbb{G} \right\}$ bracket, some simple manipulations lead, schematically, to $\left\{ \mathbb{H}, \mathbb{G} \right\} = \mathbb{G} + \mathcal{A}$ while Eq. (\ref{eq:HgGFull}) still holds\footnote{$\left\{ \mathbb{H}_\frak{m}, \mathbb{G} \right\}$ does not need to be considered in the following discussion as it is not anomalous anymore once counter-terms are set to zero, see (\ref{eq:HmGFull1}-\ref{eq:HmGFull2}).}. Therefore, similarly to $\left\{ \mathbb{H}, \mathbb{D} \right\}$, the $\left\{ \mathbb{H}, \mathbb{G} \right\}$ bracket is inherently ambiguous: multiple solutions do exist to the anomaly freedom requirement. However, explicit calculations from the previous section reveal that all the anomalies of the $\left\{ \mathbb{H}, \mathbb{G} \right\}$ bracket are \textit{equal} to some of the anomalies of the $\left\{ \mathbb{H}, \mathbb{D} \right\}$ bracket. In other words, since these anomalies must vanish to ensure the consistency of $\left\{ \mathbb{H}, \mathbb{D} \right\}$, the ambiguity of  $\left\{ \mathbb{H}, \mathbb{G} \right\}$ is fixed as well. It should be noticed that, if the diffeomorphism constraint is holonomy corrected, the ambiguity of $\left\{ \mathbb{H}, \mathbb{G} \right\}$ \textit{cannot} be fixed by the above reasoning since anomalies of $\left\{ \mathbb{H}, \mathbb{G} \right\}$ and $\left\{ \mathbb{H}, \mathbb{D} \right\}$ become independent.\\

Importantly, no ambiguity arises in the computation of the $\left\{ \mathbb{H}, \mathbb{H} \right\}$ bracket. The explicit derivation leads to  $\left\{ \mathbb{H}, \mathbb{H} \right\} \supset \mathbb{D}$, as expected from classical consideration. Still, a question remains: could we have $\left\{ \mathbb{H}, \mathbb{H} \right\} \supset \mathbb{G}$ and/or $\left\{ \mathbb{H}, \mathbb{H} \right\} \supset \mathbb{H}$, which would lead to ambiguities? This answer is negative. By construction, one cannot consistently obtain $\left\{ \mathbb{H}, \mathbb{H} \right\} \supset \mathbb{G}$ due to the peculiar summation of the Gauss constraint via the Levi-Civita tensor, see Eq.~(\ref{eq:GaussExpansionDef}), which are not recovered in the $\left\{ \mathbb{H}, \mathbb{H} \right\}$ computation. Moreover, all anomalies in the $\left\{ \mathbb{H}, \mathbb{H} \right\}$ bracket are proportional to a perturbation of the canonical variables, i.e. $\delta E$, $\delta K$, $\delta \phi$ or $\delta \pi$. Consequently, neither $\mathbb{H}_\frak{g}^{(0)}$ nor $\mathbb{H}_\frak{m}^{(0)}$ can be constructed while keeping an anomaly-free algebra. It is therefore not possible to obtain $\left\{ \mathbb{H}, \mathbb{H} \right\} \supset \mathbb{H}$. Hence, it can be concluded that no ambiguity remains for the $\left\{ \mathbb{H}, \mathbb{H} \right\}$ bracket.\\

As the ambiguities are fixed and as only a single path remains to obtain a first-class algebra of constraints, let us come back to the question of a possible closure of the algebra for scalar modes by mean of a specific GHC. As discussed above, all anomalies need to be cancelled one by one, leading to restrictions on the form of the correction. We focus on the previously derived anomalies $\mathcal{A}_1^{{\left\{ \mathbb{H}, \mathbb{D} \right\}}}$ (\ref{eq:AHD1}), $\mathcal{A}_3^{{\left\{ \mathbb{H}, \mathbb{D} \right\}}}$ (\ref{eq:AHD3}), and $\mathcal{A}_{13}^{{\left\{ \mathbb{H}, \mathbb{D} \right\}}}$ (\ref{eq:AHH13}). Without counter-terms, $\mathcal{A}_{13}^{{\left\{ \mathbb{H}, \mathbb{H} \right\}}}$ (\ref{eq:AHD13}) leads to the consistency condition:
\begin{equation}
    \tilde{g}(\frak{c}, \frak{p}) = \frak{c} + f(\frak{p}),
\end{equation}
where $f(\frak{p})$ can, at this stage, be any function that vanishes at the classical limit $\frak{p}\rightarrow \infty$. Given this solution and $\mathcal{A}_1^{{\left\{ \mathbb{H}, \mathbb{D} \right\}}}=0$ (\ref{eq:AHD1}), one obtains:
\begin{equation}
    g(\frak{c}, \frak{p}) = \pm \sqrt{\frac13 \biggl( 3 \frak{c}^2 + 4 \frak{c} f(\frak{p}) + f(\frak{p})^2 \biggr)}.
\end{equation}
Finally, $\mathcal{A}_3^{{\left\{ \mathbb{H}, \mathbb{D} \right\}}}=0$ (\ref{eq:AHD3}) implies:
\begin{equation}
    f(\frak{p}) = 0.
\end{equation}

This leads to the conclusion that $\tilde{g}(\frak{c}, \frak{p}) = \frak{c}$ and $g(\frak{c}, \frak{p})=\frak{c}$. In other words, the only way to obtain an algebra of constraints which is closed, not by the use of counter-terms but by mean of a specific choice of the GHC, is to consider the classical case without holonomy correction.

%% file: texts/conclusion.tex
In this article, we have provided an extensive material for the calculation of  Poisson brackets and anomalies in holonomy-corrected effective loop quantum cosmology. 
We tried to make all hypotheses, assumptions, and manipulations explicit so that future works can rely on a clear basis without the need for re-inventing all the heavy machinery.\\

We have taken this opportunity to provide some new insights. First, we have explained why implementing the holonomy correction in the cosmological perturbations -- either in the usual form or through a generalized function -- does {\it not} change the observational predictions of the theory. The argument is made simple and explicit.\\ 

We have also underlined that, because there are many different ways to get a fist class algebra, trying to restrict the kind of possible generalized holonomy correction so as to ensure the anomaly freedom without counter-terms is plagued with ambiguities. Even more importantly, a careful analysis shows that the algebra cannot be close without counter-terms.\\

Several points still deserve a closer look in the future:
\begin{itemize}
    \item there are theoretical arguments preventing effective (quantum) correction to the diffeomorphism constraint due to its handling from the LQG view-point  \cite{Ashtekar:1995zh} but it would still make sense at the heuristic level due to interesting results in the literature \cite{BenAchour:2016leo, Arruga:2019kyd}. Moreover, we would like to mention that it could even make sense at a deeper level \cite{Laddha:2011mk},
    \item even when implemented in the perturbative expansion of the constraints, the holonomy correction used so far (in this work and in others) is based on a modification of the background curvature through the replacement $\frak{c} \longrightarrow g(\frak{c}, \frak{p})$. An important step forward for the study of perturbations would be to consider a rigorous treatment of the perturbative expansion of the holonomy, based on the perturbed connection,
    \item counter-terms $\alpha_i$ are currently added to the deformed constraints as ``$\alpha_i\times($term entering the classical constraint$)$". This is indeed enough to close the algebra. However, following an even more general path, all the possible terms at each order of the perturbative development could be added, with the only requirement that the classical limit is recovered,
    \item it would be worth considering the possibility that the counter-terms are not only functions of the geometrical phase-space variables background parts $\frak{c}$ and $\frak{p}$, but of the full phase space variables,
    \item inverse-volume corrections should also be added, keeping the same level of generality than for the holonomy corrections. In particular, the claimed signature change should be readdressed in this framework where its occurrence is much less obvious. Some freedom about the spacetime signature at high energy is already underlined in the present work, in which only holonomy corrections were considered, through the $f$ function in Eq. (\ref{alg}),
    \item the cosmological observables should be exhaustively investigated, generalizing the results of \cite{Renevey:2021tmh,DeSousa:2022rep}. In particular, the freedom associated with the $f$ function and its consequences must be studied.
\end{itemize}

%% file: texts/appendices/summary_anomalies.tex
This appendix provides the expressions for all the anomalies calculated in Section\;(\ref{sec:AlgComputations}). The steps of the derivation can be found in the main text.\\

From the $\left\{ \mathbb{H}, \mathbb{G} \right\}$ bracket:
\begin{align}
    \mathcal{A}_1^{\left\{ \mathbb{H}, \mathbb{G} \right\}}&= g^2 + 2 \frak{c} \bigl[ \tilde{g} + \alpha_6 \bigr] - \bigl[\tilde{g}^2 + \alpha_7\bigr]+ 4 \frak{p} g \partial_\frak{p} g, \\
     \mathcal{A}_2^{\left\{ \mathbb{H}, \mathbb{G} \right\}}&=\frak{c}\bigl[ 1 + \alpha_4 \bigr]+\bigl[\tilde{g} + \alpha_6\bigr] - 2 g \partial_\frak{c} g, \\
    \mathcal{A}_3^{\left\{ \mathbb{H}, \mathbb{G} \right\}}&=-\beta_{10}, \\
    \mathcal{A}_4^{\left\{ \mathbb{H}, \mathbb{G} \right\}}&= \beta_{12}.
\end{align}
For the $\left\{ \mathbb{H}, \mathbb{D} \right\}$ bracket:
\begin{align}
     \mathcal{A}_1^{{\left\{ \mathbb{H}, \mathbb{D} \right\}}} &= 3 g^2 -  \bigl[\tilde{g}^2 + \alpha_2 \bigr] - 2 \frak{c} \bigl[ \tilde{g} + \alpha_1  \bigr], \\
    \mathcal{A}_2^{{\left\{ \mathbb{H}, \mathbb{D} \right\}}}&= \alpha_8 - \alpha_7, \\
    \mathcal{A}_3^{{\left\{ \mathbb{H}, \mathbb{D} \right\}}} &= \frak{c}\bigl[1 + \alpha_5 \bigr] + \bigl[\tilde{g} + \alpha_6 \bigr] - 2 g \partial_\frak{c}g, \\
    \mathcal{A}_4^{{\left\{ \mathbb{H}, \mathbb{D} \right\}}} &= g^2 - 2 \frak{c}\bigl[ \tilde{g} + \alpha_6 \bigr] + \bigl[\tilde{g}^2 + \alpha_7 \bigr] + 4 \frak{p} g \partial_\frak{p}g ,  \\
    \mathcal{A}_5^{{\left\{ \mathbb{H}, \mathbb{D} \right\}}} &= \frak{c}\bigl[1 + \alpha_4] + \bigl[\tilde{g}  + \alpha_6\bigr] - 2 g \partial_\frak{p}g, \\
    \mathcal{A}_6^{{\left\{ \mathbb{H}, \mathbb{D} \right\}}} &= \beta_5 - \beta_6, \\
     \mathcal{A}_7^{{\left\{ \mathbb{H}, \mathbb{D} \right\}}} &= 2 \beta_1 - \beta_2, \\
      \mathcal{A}_8^{{\left\{ \mathbb{H}, \mathbb{D} \right\}}} &= \beta_4, \\
      \mathcal{A}_9^{{\left\{ \mathbb{H}, \mathbb{D} \right\}}} &= - \beta_{10}, \\
      \mathcal{A}_{10}^{{\left\{ \mathbb{H}, \mathbb{D} \right\}}} &= \beta_{12}, \\
        \mathcal{A}_{11}^{{\left\{ \mathbb{H}, \mathbb{D} \right\}}} &= \beta_{10} - 2 \beta_6 + \beta_9, \\
        \mathcal{A}_{12}^{{\left\{ \mathbb{H}, \mathbb{D} \right\}}} &= \beta_8, \\
        \mathcal{A}_{13}^{{\left\{ \mathbb{H}, \mathbb{D} \right\}}} &= \beta_{11} - \beta_{12}.
\end{align}
For the $\left\{ \mathbb{H}, \mathbb{H} \right\}$ bracket:
\begin{align}
    \mathcal{A}_1^{{\left\{ \mathbb{H}, \mathbb{H} \right\}}} &= 2\bigl[ \tilde{g} + \alpha_1 \bigr] \biggl( g \ppc{g} - \tilde{g} - \alpha_6 \biggr)  - \biggl( g^2 + 4 \frak{p} g \ppp{g}\biggr) \pc{\tilde{g}+\alpha_1}   \notag \\
    &\qquad+ 4 \frak{p} g \ppc{g} \pp{\tilde{g}+\alpha_1} + \frac12 \bigl[ \tilde{g}^2 + \alpha_2 \bigr] \biggl( 2 + 3\alpha_5 - \alpha_4 \biggr),\\
     \mathcal{A}_2^{{\left\{ \mathbb{H}, \mathbb{H} \right\}}} &= \bigl[ \tilde{g}^2 + \alpha_2 \bigr] \biggl( \tilde{g} + \alpha_6 - g \ppc{g} \biggr) - \frac12 \biggl( g^2 + 4 \frak{p} g \ppp{g} \biggr) \pc{\tilde{g}^2+\alpha_2}  \notag\\
      &\qquad+ 2 \frak{p} g \ppc{g} \pp{\tilde{g}^2 + \alpha_2} + \bigl[ \tilde{g}+\alpha_1\bigr]\biggl( \tilde{g}^2 + 3 \alpha_8 - 2 \alpha_7\biggr), \\
      \mathcal{A}_3^{{\left\{ \mathbb{H}, \mathbb{H} \right\}}} &= \bigl[ 1 + \alpha_3 \bigr] \bigl[ \alpha_5 - \alpha_4 \bigr], \\
       \mathcal{A}_4^{{\left\{ \mathbb{H}, \mathbb{H} \right\}}} &= \bigl[ 1 + \alpha_3 \bigr] \biggl(g\ppc{g} + \bigl[ \tilde{g} + \alpha_6 \bigr] + k \bigl[ 1 + \alpha_5 \bigr] \biggr) - \bigl[ \tilde{g} + \alpha_1 \bigr] \bigl[ 1 + \alpha_9 \bigr] + \notag \\
       &\qquad+ 2 \frak{p} g \ppc{g} \ppp{\alpha_3} + \biggl( \frac12 g^2 + 2 \frak{p} \ppp{g} \biggr) \ppc{\alpha_3}., \\
       \mathcal{A}_5^{{\left\{ \mathbb{H}, \mathbb{H} \right\}}} &= \partial_\frak{c}\beta_1,\\
    \mathcal{A}_6^{{\left\{ \mathbb{H}, \mathbb{H} \right\}}} &= \partial_\frak{c}\beta_3,\\
    \mathcal{A}_7^{{\left\{ \mathbb{H}, \mathbb{H} \right\}}} &= -\partial_\frak{c}\beta_2,\\
    \mathcal{A}_8^{{\left\{ \mathbb{H}, \mathbb{H} \right\}}} &= -\partial_\frak{c}\beta_4,\\
    \mathcal{A}_9^{{\left\{ \mathbb{H}, \mathbb{H} \right\}}} &= \pc{\beta_2 + \beta_4}, \\
     \mathcal{A}_{10}^{{\left\{ \mathbb{H}, \mathbb{H} \right\}}} &= \beta_1 - \beta_3 - \beta_5 - \beta_3 \beta_5,\\
    \mathcal{A}_{11}^{{\left\{ \mathbb{H}, \mathbb{H} \right\}}} &= \beta_1 + \beta_7 + \beta_1 \beta_7 - \beta_3, \\
    \mathcal{A}_{12}^{{\left\{ \mathbb{H}, \mathbb{H} \right\}}} &= \beta_1 + \beta_3 + \beta_6 + \beta_3 \beta_6 + \beta_8 + \beta_1 \beta_8 - \beta_2 - \beta_4, \\
    \mathcal{A}_{13}^{{\left\{ \mathbb{H}, \mathbb{H} \right\}}} &=  \bigl[ 1 + \beta_2 \bigr]\bigl( 2 - \alpha_4 + 3 \alpha_5 \bigr) - 2 \pc{\tilde{g}+ \alpha_1}, \\
    \mathcal{A}_{14}^{{\left\{ \mathbb{H}, \mathbb{H} \right\}}} &= \bigl[ 1 + \beta_4 \bigr]\bigl( -2 + \alpha_4 - 3 \alpha_5 \bigr) + 2 \pc{\tilde{g}+ \alpha_1}, \\
    \mathcal{A}_{15}^{{\left\{ \mathbb{H}, \mathbb{H} \right\}}} &= \ppc{\alpha_3}, \\
    \mathcal{A}_{16}^{{\left\{ \mathbb{H}, \mathbb{H} \right\}}} &= - 6\bigl[ \tilde{g}+ \alpha_1 \bigr] \bigl[ 1 + \beta_6 \bigr] + 2g \ppc{g} \bigl( 3 + 3 \beta_1 - 2 \frak{p} \ppp{\beta_1} \bigr) + g^2 \ppc{\beta_1} + 4 g \frak{p} \ppp{g} \ppc{\beta_1}, \\
    \mathcal{A}_{17}^{{\left\{ \mathbb{H}, \mathbb{H} \right\}}} &= 12 \tilde{g} + 10 \alpha_1 + 2 \alpha_6 + 4 \bigl[\tilde{g}+\alpha_1\bigr] \beta_{10} + 2 \bigl[\tilde{g}+\alpha_6\bigr] \beta_2  + 6 \tilde{g} \beta_9 - 2 \tilde{g} \ppc{\tilde{g}}  \notag\\ &\qquad - 2 g \ppc{g} \bigl( 5 + 5 \beta_2 - 2 \frak{p} \ppp{\beta_2} \bigr)  - \ppc{\alpha_2} - g^2 \ppc{\beta_2} - 4 g \frak{p} \ppp{g} \ppc{\beta_2},\\
    \mathcal{A}_{18}^{{\left\{ \mathbb{H}, \mathbb{H} \right\}}} &= 2 \alpha_1 - 2 \alpha_6 + 6 \bigl[\tilde{g} + \alpha_1 \bigr] \beta_{11} - 4 \bigl[\tilde{g}+\alpha_1\bigr] \beta_{12} - 2 \bigl[\tilde{g} + \alpha_6 \bigr]\beta_4 + 2 \tilde{g} \ppc{\tilde{g}}   \notag \\
    &\qquad - 2 g \ppc{g} \bigl( 1 + \beta_4 + 2 \frak{p} \ppp{\beta_4} \bigr) + \ppc{\alpha_2} + g^2 \ppc{\beta_4} + 4 g \frak{p} \ppp{g} \ppc{\beta_4}, \\
    \mathcal{A}_{19}^{{\left\{ \mathbb{H}, \mathbb{H} \right\}}} &= 3 \frak{p} \bigl[ \tilde{g} + \alpha_1 \bigr] + 3 \frak{p} \bigl[\tilde{g}+\alpha_1\bigr] \beta_8- g \ppc{g} \bigl( 3 \frak{p} \bigl[ 1 + \beta_3 \bigr] + 2 \frak{p}^2 \ppp{\beta_3} \bigr) + \frac{\frak{p}}{2} g^2 \ppc{\beta_3}  \notag\\
    &\qquad  + 2 \frak{p}^2 g \ppp{g} \ppc{\beta_3}.
\end{align}

%% file: texts/appendices/anomalies_table_correspondance.tex
Some notations differ between previous works, such as \cite{Han:2017wmt, Cailleteau:2011kr}, and this study. To help the unfamiliar reader and to allow an easy comparison between approaches\footnote{Note that some anomalies are equals modulo the sign which can be changed by an integration by parts given the null boundary conditions.}{$^,$}\footnote{Note that in \cite{Han:2017wmt}, there is a misprint in Eq. $(52)$: anomalies $\mathcal{A}_{32}$ and $\mathcal{A}_{33}$ are, respectively, presented as $\mathcal{A}_{33}$ and $\mathcal{A}_{34}$ in Eqs. (53) and (54).}, we provide in this appendix a table of correspondence for the anomalies (N.A. stands for ``non available").

\begin{longtable}{| c|c||c|c|c |} 
\caption{$\left\{ \mathbb{H}, \mathbb{G} \right\}$ bracket - Table of correspondence}\\
\hline
Bracket& Sub-bracket &This work & Han \& Liu's work \cite{Han:2017wmt} & Cailleteau \& al's work \cite{Cailleteau:2011kr}\\
\hline 
\multirow{8}{3em}{$\left\{ \mathbb{H}, \mathbb{G} \right\}$} & \multirow{4}{3.5em}{$\left\{ \mathbb{H}_\frak{g}, \mathbb{G} \right\}$} & & &\\
&& $\mathcal{A}_1^{\left\{ \mathbb{H}, \mathbb{G} \right\}}$ & $\mathcal{A}_{34}$& N.A.\\ 
&& $\mathcal{A}_2^{\left\{ \mathbb{H}, \mathbb{G} \right\}}$ & $\mathcal{A}_{33}$ & N.A.\\
&&&&\\
\cline{2-5}
& \multirow{4}{3.5em}{$\left\{ \mathbb{H}_\frak{m}, \mathbb{G} \right\}$} &&&\\
& & $\mathcal{A}_3^{\left\{ \mathbb{H}, \mathbb{G} \right\}}$& N.A.&N.A.\\
& & $\mathcal{A}_4^{\left\{ \mathbb{H}, \mathbb{G} \right\}}$& N.A.&N.A.\\
& & & &\\
\hline
\end{longtable}

\begin{longtable}{| c|c||c|c|c |} 
\caption{$\left\{ \mathbb{H}, \mathbb{D} \right\}$ bracket - Table of correspondence}\\
\hline
Bracket& Sub-bracket &This work & Han \& Liu's work \cite{Han:2017wmt} & Cailleteau \& al's work \cite{Cailleteau:2011kr}\\
\hline
\multirow{17}{3em}{$\left\{ \mathbb{H}, \mathbb{D} \right\}$}& \multirow{7}{3.5em}{$\left\{ \mathbb{H}_\frak{g}, \mathbb{D}_\frak{g} \right\}$} & & &\\
& & $\mathcal{A}_1^{\left\{ \mathbb{H}, \mathbb{D} \right\}}$ & $\mathcal{A}_1$ & $\mathcal{A}_1$ \\
&& $\mathcal{A}_2^{\left\{ \mathbb{H}, \mathbb{D} \right\}}$ & $\mathcal{A}_5$ & $\mathcal{A}_4$ \\
&& $\mathcal{A}_3^{\left\{ \mathbb{H}, \mathbb{D} \right\}}$ & $\mathcal{A}_2$ & $\mathcal{B}$ \\
&& $\mathcal{A}_4^{\left\{ \mathbb{H}, \mathbb{D} \right\}}$ & $\mathcal{A}_4$ & $\mathcal{A}_3$ \\
&& $\mathcal{A}_5^{\left\{ \mathbb{H}, \mathbb{D} \right\}}$ & $\mathcal{A}_3$ & $\mathcal{A}_2$ \\
&& & &\\
\cline{2-5}
& \multirow{10}{3.5em}{$\left\{ \mathbb{H}_\frak{m}, \mathbb{D}\right\}$}& & &\\
&& $\mathcal{A}_6^{\left\{ \mathbb{H}, \mathbb{D} \right\}}$ & $\mathcal{A}_9$ & N.A\footnote{In \cite{Cailleteau:2011kr}, no deformation of the matter sector were considered, thus no anomalies related \textit{only} to those counter-terms are found.} \\
 && $\mathcal{A}_7^{\left\{ \mathbb{H}, \mathbb{D} \right\}}$& $\mathcal{A}_7$& N.A\\
 && $\mathcal{A}_8^{\left\{ \mathbb{H}, \mathbb{D} \right\}}$& $\mathcal{A}_6$& N.A \\
 && $\mathcal{A}_9^{\left\{ \mathbb{H}, \mathbb{D} \right\}}$& $\mathcal{A}_{12}$& N.A\\
 && $\mathcal{A}_{10}^{\left\{ \mathbb{H}, \mathbb{D} \right\}}$& $\mathcal{A}_{10}$& N.A\\
 && $\mathcal{A}_{11}^{\left\{ \mathbb{H}, \mathbb{D} \right\}}$& $\mathcal{A}_{13}$&N.A\\
 && $\mathcal{A}_{12}^{\left\{ \mathbb{H}, \mathbb{D} \right\}}$& $\mathcal{A}_{8}$&N.A\\
 && $\mathcal{A}_{13}^{{\left\{ \mathbb{H}, \mathbb{D} \right\}}}$& $\mathcal{A}_{11}$&N.A\\
& && &\\
\hline
\end{longtable}

\begin{longtable}{| c|c||c|c|c |} 
\caption{$\left\{ \mathbb{H}, \mathbb{H} \right\}$ bracket - Table of correspondence}\\
\hline
Bracket& Sub-bracket &This work & Han \& Liu's work \cite{Han:2017wmt} & Cailleteau \& al's work \cite{Cailleteau:2011kr}\\
\hline
\multirow{25}{3em}{$\left\{ \mathbb{H}, \mathbb{H} \right\}$}& \multirow{6}{3.5em}{$\left\{ \mathbb{H}_\frak{g}, \mathbb{H}_\frak{g} \right\}$}& & &\\
&& $\mathcal{A}_1^{\left\{ \mathbb{H}, \mathbb{H} \right\}}$ & $\mathcal{A}_{16}$ & $\mathcal{A}_7$ \\
&& $\mathcal{A}_2^{\left\{ \mathbb{H}, \mathbb{H} \right\}}$ & $\mathcal{A}_{17}$ & $\mathcal{A}_8$ \\
&& $\mathcal{A}_3^{\left\{ \mathbb{H}, \mathbb{H} \right\}}$ & $\mathcal{A}_{14}$ & $\mathcal{A}_5$ \\
&& $\mathcal{A}_4^{\left\{ \mathbb{H}, \mathbb{H} \right\}}$ & $\mathcal{A}_{15}$ & $\mathcal{A}_6$ \\
&& & &\\
\cline{2-5}
&\multirow{10}{4em}{$\left\{ \mathbb{H}_\frak{m}, \mathbb{H}_\frak{m} \right\}$}& & &\\
&& $\mathcal{A}_5^{\left\{ \mathbb{H}, \mathbb{H} \right\}}$ & $\mathcal{A}_{31}$ & N.A \\
&& $\mathcal{A}_6^{\left\{ \mathbb{H}, \mathbb{H} \right\}}$ & $\mathcal{A}_{30}$ & N.A \\
&& $\mathcal{A}_7^{\left\{ \mathbb{H}, \mathbb{H} \right\}}$ & $\mathcal{A}_{28}$ & N.A \\
&& $\mathcal{A}_8^{\left\{ \mathbb{H}, \mathbb{H} \right\}}$ & $\mathcal{A}_{25}$ & N.A \\
&& $\mathcal{A}_9^{\left\{ \mathbb{H}, \mathbb{H} \right\}}$ & $\mathcal{A}_{27}$ & N.A \\
&& $\mathcal{A}_{10}^{\left\{ \mathbb{H}, \mathbb{H} \right\}}$ & $\mathcal{A}_{32}$ & N.A \\
&& $\mathcal{A}_{11}^{\left\{ \mathbb{H}, \mathbb{H} \right\}}$ & $\mathcal{A}_{29}$ & N.A \\
&& $\mathcal{A}_{12}^{\left\{ \mathbb{H}, \mathbb{H} \right\}}$ & $\mathcal{A}_{26}$ & N.A \\
&& & &\\
\cline{2-5}
&\multirow{9}{4em}{$\left\{ \mathbb{H}_\frak{g}, \mathbb{H}_\frak{m} \right\}$}& & &\\
&& $\mathcal{A}_{13}^{\left\{ \mathbb{H}, \mathbb{H} \right\}}$ & $\mathcal{A}_{19}$ & N.A \\
&& $\mathcal{A}_{14}^{\left\{ \mathbb{H}, \mathbb{H} \right\}}$ & $\mathcal{A}_{20}$ & N.A \\
&& $\mathcal{A}_{15}^{\left\{ \mathbb{H}, \mathbb{H} \right\}}$ & $\mathcal{A}_{18}$ & $\mathcal{A}_{9}$ \\
&& $\mathcal{A}_{16}^{\left\{ \mathbb{H}, \mathbb{H} \right\}}$ & $\mathcal{A}_{23}$ & N.A \\
&& $\mathcal{A}_{17}^{\left\{ \mathbb{H}, \mathbb{H} \right\}}$ & $\mathcal{A}_{21}$ & $\mathcal{A}_{12}$ \\
&& $\mathcal{A}_{18}^{\left\{ \mathbb{H}, \mathbb{H} \right\}}$ & $\mathcal{A}_{22}$ & $\mathcal{A}_{13}$ \\
&& $\mathcal{A}_{19}^{\left\{ \mathbb{H}, \mathbb{H} \right\}}$ & $\mathcal{A}_{24}$ & N.A \\
&& & &\\
\hline
\end{longtable}